\documentclass[aps,preprint,superscriptaddress,nofootinbib,preprintnumbers]{revtex4-1}

\usepackage{amsmath,amssymb,amsfonts}
\usepackage{bm, slashed, soul}
\usepackage{color,graphicx}
\usepackage{hyperref}

\usepackage{xcolor}
\definecolor{red}{rgb}{0.9, 0,0}

\def\tr{{\rm Tr}}
\def\tilq{\tilde{q}}

\def\FO{{{\cal F}_{\cal O}}}

\newcommand{\be}{\begin{equation}}
\newcommand{\ee}{\end{equation}}
\newcommand{\bea}{\begin{eqnarray}}
\newcommand{\eea}{\end{eqnarray}}

\def\beq#1\eeq{\begin{align}#1\end{align}}
\def\beqnn#1\eeq{\begin{align*}#1\end{align*}}
\newcommand{\nn}{\nonumber}

\newcommand{\frud}[2]{\left( \frac{#1}{{#2}}\right)}

\def \tev {\text{ TeV}}
\def \gev {\text{ GeV}}

\def\mdm{m_{DM}}

\begin{document}

\hfill

\vspace{1.5cm}

\begin{center}
{\Huge\bf
Twin Baryogenesis
}
\\ \vspace*{0.3cm}

\bigskip\vspace{0.6cm}
{{\bf Marco Farina, Angelo Monteux and Chang Sub Shin \footnote{Email:~\url{farina.phys@gmail.com},~\url{amonteux@physics.rutgers.edu},~\url{changsub@physics.rutgers.edu}}}} \\[5mm]
{\it   New High Energy Theory Center, Department of Physics, Rutgers University,\\136 Frelinghuysen Road, Piscataway, NJ 08854, USA}\\[3mm]
\end{center}
\vskip.5cm
\baselineskip 15pt
\bigskip
\centerline{\large\bf Abstract}

\begin{quote}
In the context of Twin Higgs models, we study a simple mechanism that simultaneously generates asymmetries in the dark and visible sector through the out-of-equilibrium decay of a TeV scale particle charged under a combination of baryon and twin baryon number. We predict the dark matter to be a 5 GeV twin baryon, which is easy to achieve because of the similarity between the two confinement scales. Dark matter is metastable and can decay to three quarks, yielding indirect detection signatures. The mechanism requires the introduction of a new colored particle, typically within the reach of the LHC, of which we study the rich collider phenomenology, including prompt and displaced dijets, multi-jets, monojets and monotops. 
\end{quote}

\thispagestyle{empty}

 \newpage
 
\setcounter{page}{1}

\section{Introduction}

Understanding the nature of dark matter and the origin of its abundance is a long standing problem, in the sense that we do not yet have enough observational data to distinguish alternatives. One remarkable observation is  the similarity between the measured abundances of baryons and of dark matter, $\Omega_{\text{DM}} \simeq 5\, \Omega_B  $~\cite{Ade:2015xua}. In many models describing the early universe, this coincidence is just an accident and these two quantities are unrelated, and obtained from different production mechanisms. However, such coincidence could be an indication of a common origin. 
A notable class of models addressing this possibility falls under the category of Asymmetric Dark Matter (ADM)~\cite{Nussinov:1985xr,Chivukula:1989qb,Barr:1990ca,Kaplan:1991ah,Hooper:2004dc,Kaplan:2009ag} (for recent reviews see~\cite{Davoudiasl:2012uw,Petraki:2013wwa,Zurek:2013wia}): here, dark matter is charged under a new global conserved $U(1)_X$ and, similarly to what happens for the baryons, an asymmetry is generated between particles and anti-particles in the dark matter sector. The symmetric component is assumed to annihilate efficiently at a later time, after which the dark matter abundance is given by the remaining asymmetric component.

A common mechanism can be responsible for the generation of both asymmetries, naturally yielding $n_{DM}\sim n_B$ (where $n_i$ is the number density of the particle species $i$).
This is usually done by assuming that an asymmetry is
initially generated in either one (or both) sectors, and then some $B$- (or $B\!-\!L$ if above the electroweak scale) and $X$-violating operator in thermal equilibrium transfers the baryon asymmetry into the dark matter sector, or vice versa. Once the transfer operator falls out of equilibrium, the (asymmetric) number densities $n_B$ and $n_{DM}$ freeze out and their ratio gives a prediction for the dark matter mass $m_{DM}$. Indeed from
\begin{equation}
  \frac{\Omega_{DM}}{\Omega_B} = \frac{n_{DM}}{n_B} \frac{m_{DM}}{m_B} \,\, ,
\end{equation}
the coincidence $\Omega_{DM}\simeq 5\,\Omega_B$ becomes $m_{{DM}}\sim 5 \, m_B $. It is therefore clear that ADM does not completely resolve the puzzle of coincidence of densities, and it actually translates it in a coincidence of masses. If baryons and DM are expected to belong to two different sectors, their masses would be unrelated, then why should they be so close to each other?
A natural answer is to suppose DM to be composed of dark baryons, with  $U(1)_X$ identified by dark baryon number and mass generated by a copy of QCD. Early realizations exploited this idea in the context of mirror world (see \cite{Berezhiani:2005ek} for a review).  Twin Higgs~\cite{Chacko:2005pe,Chacko:2005un,Chacko:2005vw,Burdman:2014zta,Barbieri:2005ri} models provide another natural framework to implement asymmetric dark matter and explain the coincidence of masses. This is tantalizing as Twin Higgs was originally introduced  as a solution to the little Hierarchy problem, maintaining naturalness of a light Higgs boson without introducing new colored degrees of freedom (the SM Higgs is a pseudo-Goldstone boson and is protected from quadratic divergences). In  the Twin Higgs mechanism the field content of the Standard Model (SM) is doubled, with additional ``twin'' quarks, leptons and gauge groups, resulting in a mirror sector that could very well include a stable dark matter candidate. This has been explored in recent works, e.g.~\cite{Garcia:2015toa,Farina:2015uea}, where it has been shown that twin baryons are a viable dark matter candidate. These works focused on the DM properties while the origin of the asymmetry itself was left as a UV problem. This approach is especially justified for Twin Higgs, as the theory requires a UV completion at a scale of $5-10\tev$ and any high-energy baryogenesis mechanism will depend on the particular UV interactions.

Nevertheless, the baryon asymmetry might also be generated at low temperatures, well below the Twin Higgs UV cutoff. 
Then, we can explicitly calculate the asymmetries and all the observational consequences from the model. So far, this has not been explored. 
In this paper, we will describe a mechanism in which the baryon and dark asymmetries are simultaneously generated by the decay of a new Dirac singlet with a mass of ${\cal O}$(TeV) at low temperatures.
The field content is similar to (twice) the ``WIMP baryogenesis'' scenarios~\cite{Cui:2012jh,RompineveSorbello:2013xwa}, with the crucial difference being the presence of Dirac fermions instead of Majorana. In the latter sense it is related to hylogenesis \cite{Davoudiasl:2010am} with the main differences being the complete specification of the model via renormalizable operators and the composite nature of dark matter.

The model breaks ordinary and twin baryon numbers ($B$ and $\tilde B$) but conserves $B \!-\!\tilde B$, such that an equal number of baryons and twin baryons populate the present universe, $n_B=n_{DM}$. 
The baryon asymmetry is generated by diagrams with twin quarks in the loop, and vice versa. Depending on the mass spectrum, the asymmetry can be generated at either one or two loops. 
The ratio of baryon and dark matter abundances predicts $m_{DM}\simeq 5\gev$, which can be naturally realized in Twin Higgs models. While in this work we focus on the ``vanilla'' Twin Higgs model with three twin generations, the mechanism works equally well for the ``fraternal Twin Higgs''~\cite{Craig:2015pha}. As a matter of fact, the mechanism is general and needs not to be related to solutions of the little hierarchy problem, although it particularly compelling that it is possible to explain baryogenesis, dark matter and naturalness of the weak scale at once.

The cosmological and phenomenological consequences of our model are also fully calculable.
 Because of $B$ and $\tilde B$ violation,  dark matter is unstable and its lifetime lies in an interesting region of parameter space, around the lower bounds found from DM indirect detection experiments. In addition to the Dirac fermions, we need to introduce at least one colored scalar with a mass of ${\cal O}(100\,{\rm GeV}-{\rm TeV})$ in each sector, which results in multiple interesting signatures at the LHC. As such, the model presented in this paper is placed at the intersection of independent experimental probes, each pushing in complementary directions. Still, we find large viable regions of parameter space.

This paper is organized in the following way: in Section~\ref{sec:TH} we review the Twin Higgs mechanism and briefly discuss the candidates for dark matter in the twin sector. In Section~\ref{sec:BG} we introduce the new degrees of freedom and compute the baryon and dark matter asymmetries, and also detail the thermal history of the universe leading to baryogenesis. We discuss the bounds on the unstable dark matter candidate in Section~\ref{sec:DMind} and the relevant LHC searches in Section~\ref{sec:ColliderPheno}; here we also show the complementarity of the different experimental signatures and show detailed results for two choices for universal and hierarchical parameters. We conclude in Section~\ref{concl} with an outlook. 


\section{Twin Higgs Framework and Dark Matter}\label{sec:TH}
We start by briefly describing the Twin Higgs mechanism \cite{Chacko:2005pe,Chacko:2005un}, which implements the Higgs as a pseudo Goldstone boson, solving the little hierarchy problem with new uncolored particles. A complete copy of the SM content is added, including gauge symmetries\footnote{Whether hypercharge is gauged or not is left to model building (e.g. recent version of composite Twin Higgs with no gauging \cite{Barbieri:2015lqa}), and even if present the twin photon is expected to be either massless or at $m_{\tilde \gamma} \simeq 4 \pi f$. Our main results do not depend on the nature of twin hypercharge and we leave comments on cosmological constraints from $N_{\rm eff}$ to Appendix B.} $SU(3) \times SU(2)$,
 with a $Z_2$ exchange symmetry between the two copies. The Higgs doublet and its copy are taken to form a fundamental of $SU(4)$ which is assumed to be an accidental global symmetry of the Higgs potential. The global symmetry is broken to $SU(3)$ producing 7 Goldstone bosons, 6 of which represent the longitudinal degrees of freedom of $W,Z,\tilde W,\tilde Z$; the remaining one can be identified with the SM Higgs boson and it is naturally light, as the $SU(4)$ symmetry protects its potential from quadratic contributions (here and in the following, we denote fields and couplings in the twin sector with a tilde, e.g. $\tilq$ is a twin quark).
On the other hand, a small breaking of the $Z_2$ symmetry must be introduced to give a realistic theory with $v<f$ where $v$ is the SM Higgs vacuum value and $f$ is the scale of breaking of the global symmetry. We expect the theory to be UV completed at a scale $\sim 4 \pi f$ and so for naturalness we expect $f$ of order TeV.  The twin sector is then comprised of particles with masses larger than their SM counterparts by a factor $\sim f/v$ and interactions of the SM-like Higgs of the form
\begin{equation}
  \mathcal{L} \supset y_\psi h \bar{\psi} \psi - \frac{y_\psi}{\sqrt{2}f} h^\dagger h  \bar{\tilde{\psi}} \tilde{\psi} \,\, .
\end{equation}
It is evident how the structure of the previous equation enforces cancellation of quadratic divergences. Apart from irrelevant operators this is the only interaction between the two copies involving light degrees of freedom. 
The value of $f$ is bounded from below due to the effect of modification of the Higgs couplings, so that $f\gtrsim 3v$.
As already mentioned, $Z_2$ is expected to not be fully respected and small breakings in the form of differences of couplings between the two sectors are tolerable. They however reintroduce a quadratic sensitivity to the cutoff, as such naturalness requires $y_t \simeq y_{\tilde{t}}$  at percent level and both $\tilde g_2$ and $\tilde g_3$ to be within $10\%$ of their SM counterpart~\cite{Craig:2015pha}. Given the smallness of the relative Yukawas, there are no bounds on $Z_2$ breaking for the first two generations. Indeed the first two generations of twin fermions can be omitted completely from the spectrum (referred to as the fraternal Twin Higgs~\cite{Craig:2015pha}). 
Most of our results are general and apply to both to identical and fraternal TH. 
 
In both cases the lightest baryon, either a twin neutron/proton or a twin $b$ baryon ($\tilde b\tilde b\tilde b$), is stable if twin baryon number is conserved and is expected to have a lifetime on cosmological scales if only $B-\tilde B$ is conserved. The latter is the case we are about to study, and detailed discussion of the lifetime constraints will be presented in Section~\ref{sec:DMind}. The twin baryon mass is likely linked to $\tilde \Lambda_{QCD}$ and so falls naturally in the few GeV range. A larger $\tilde \Lambda_{QCD}$ naturally arises from heavier twin quarks ($\tilde t$ and $\tilde b$) causing a steepening in the RGE running of $\tilde g_3$ between $f$ and $v$; a ten percent level mismatch between $g_3$ and $\tilde g_3$ can also significantly raise the twin confinement scale, see e.g.~\cite{Craig:2015pha,Garcia:2015toa,Farina:2015uea,Freytsis:2016dgf}.

In the following, we present a minimal baryogenesis scenario having in mind that it should be implemented in the necessary UV completion of the Twin Higgs mechanism that is expected to be present at the TeV scale (for example composite twin higgs in the composite holographic Higgs framework~\cite{Geller:2014kta,Barbieri:2015lqa,Low:2015nqa}). In this sense the choice of having  additional particles with masses around the TeV is natural.

Since we consider a low temperature baryogenesis, below the electroweak phase transition, 
the twin leptons do not play any role (this is also true for the hierarchy problem). However the spectrum of light twin leptons is cosmologically important, and will be discussed in Appendix~\ref{DR}.


\section{Setup and Asymmetry generation }\label{sec:BG}
In addition to the Twin Higgs field content, we add neutral Dirac fermions, $\Psi_A = (\psi_{A},\, i\sigma_2\tilde \psi^*_{A})^T $ and scalars $\phi,\tilde \phi$ (respectively charged under the SM color and twin color gauge groups), with the following renormalizable interactions  
\beq
\Delta {\cal L}  =&\,  M_A \psi_{A} \tilde\psi_{A} + \frac{1}{2}m_\phi^2\phi\phi^*  + 
\frac{1}{2}  m_{\tilde \phi}^2\tilde\phi\tilde\phi^*  \nonumber\\ 
& + \kappa_{A; i}\psi_{A}q^i \phi^* 
 +\lambda_{jk} \phi q^j q^k 
\nonumber \\
& + \tilde\kappa_{A; i} \tilde\psi_{A} \tilde q^i \tilde\phi^*  +  \tilde\lambda_{jk} \tilde\phi \tilde q^j \tilde q^k + h.c., 
\label{eq:RenLagrangian}
\eeq where $q^i$ ($\tilde q^i$) are the (twin) quarks and $i,j,k=1,2,3$ are flavor indices. The Dirac fermions are labeled by $A=1,2,\ldots$ and without loss of generality we consider only two species in the following. It is clear from the structure of the interactions that $\phi$ can either be up-like ($Q_\phi=2/3$) or down-like ($Q_\phi=-1/3$), leaving its coupling with two quarks to be respectively $\phi d^j d^k$ or $\phi u^j d^k$. In the first case $\lambda_{jk}$ is antisymmetric under the exchange $j\leftrightarrow k$. 

In general we expect each coupling in the SM and Twin sector to be of the same order of magnitude, which also avoids introducing (radiative) $Z_2$ breaking. 
We expect this to be particularly true for third generation quarks as the top quark and twin top control the fine-tuning of the Higgs mass. In the following, we take the $Z_2$ invariant assumption of $\kappa_A=\tilde\kappa_A$, $\lambda=\tilde\lambda$
, and
$m_\phi= m_{\tilde\phi}$. While not necessary, the new states are expected to be part of the UV completion and thus to have masses at the TeV scale.

Notice that while  $U(1)_{B}$ and $U(1)_{\tilde B}$ are individually explicitly broken in Eq.~(\ref{eq:RenLagrangian}), the combination  $U(1)_{B-\tilde B}$ is conserved by imposing charges as   
$Q_{B-\tilde B}(\tilde \phi)=- Q_{B-\tilde B}(\phi)=2/3$ and $Q_{B-\tilde B}(\Psi_A)=-1$. 
 As mentioned in the Introduction this implies that if an asymmetry for the baryon yield $(Y_{\Delta B}=({n_B-n_{\bar B}})/s)$ is produced, the same asymmetry for the twin baryons $(Y_{\Delta\tilde B})$ is generated,  
 $Y_{\Delta B}=Y_{\Delta\tilde B}$. 
The observed baryon and dark matter abundances give a sharp prediction for the dark matter mass, 
which can  be translated into predictions for the twin confining scale or twin Yukawa couplings.
The corresponding phenomenology is distinctly specified, as opposed to models that predict order-of-magnitude equalities between baryon and dark matter abundances.

We note that without the Dirac mass terms $M_A$, individual conservation of $B$ and $\tilde B$ is restored. Therefore, washout effects between the two sectors can be neglected if the asymmetries are produced well below the singlets mass, in the out-of-equilibrium decay of $\Psi$: the individual asymmetries are
$Y_{\Delta B} = Y_{\Delta\tilde B}= \epsilon_{B} Y_\Psi$, where 
$\epsilon_B$ is  the baryon asymmetry parameter, defined as 
\begin{equation}
\epsilon_{B}=\sum_f Q_B(f)({\rm Br}(\Psi \to f) -{\rm Br}(\bar\Psi\to\bar f)) \,\, ,
\end{equation}
for final states $f$ with baryon numbers $Q_B(f)$.

Since the effective cut-off of the theory is around the TeV scale, baryogenesis happens at relatively low temperatures. To be more specific and in order to calculate the actual value of $Y_{\Delta B}$, we need to describe the thermal history of the Universe leading to baryogenesis. Let us  now discuss the available possibilities.

\subsection{Thermal History}\label{thermalH}
In general $\Psi$ can be produced in two ways, thermally or non-thermally,  depending on the reheating temperature, $T_{\rm reh}$. We start with the simplest case of thermal production of $\Psi$ with sufficiently $T_{\rm reh} \gg m_A$, so that  at some high temperature, $\Psi$'s are produced and in equilibrium with the thermal bath. In analogy to a metastable WIMP, their yield prior to decay is determined by freeze-out and it is maximized in the case of relativistic decoupling. In such a case, we estimate $Y_\Psi \lesssim Y_{eq}={135 \zeta(3)}/{4\pi^4g_*(M_\Psi)} \lesssim 10^{-3} $, where $g_*(T)$ is the number of relativistic degrees of freedom at the temperature $T$. 
Reproducing the observed baryon yield, $Y_{\Delta B} \simeq 10^{-10}$, translates into the requirement $\epsilon_B\gtrsim 10^{-7}$. In this case out of equilibrium decay is required in order to meet Sakharov conditions. This corresponds to decay happening after chemical decoupling of $\Psi$, which generically happens at temperature $\sim M_A$, giving the condition $\Gamma_\Psi \ll  H(M_A)$.

Once produced, possible processes of the form $\phi \bar q \leftrightarrow \tilde \phi^* \tilde q$ and $ \bar q\bar q\bar q \leftrightarrow \tilde{q}\tilde{q}\tilde{q}$ can wash out the baryon asymmetry. The latter are higher order and so usually suppressed enough to be negligible, the former are Boltzmann-suppressed once the temperature drops below $\sim m_\phi/25$. 
If the $\Psi$ decay happens well below this temperature, the final baryon and twin baryon abundances are nearly preserved and washout effects can be neglected \cite{Cui:2012jh}.

All the previous considerations are independent of the Twin sector field content. Other constraints could apply in specific cases, for instance new (nearly) massless states could contribute to $N_{\rm eff}$ or the annihilation of other possible relics and the symmetric component of DM could be inefficient. We refer the reader to the existing literature~\cite{Garcia:2015loa,Garcia:2015toa,Craig:2015xla,Farina:2015uea,Freytsis:2016dgf}.
Let us mention that we can avoid most of these bounds and relax part of the assumptions on the new sector by non-thermally producing the $\Psi$'s via  the decay of a ``reheaton'' and a low reheating temperature, such that the twin sector is not efficiently reheated \cite{Berezhiani:1995am}. Similar ideas have appeared recently in \cite{Reece:2015lch,Adshead:2016xxj}.
Because the SM and Twin sectors are kept in thermal equilibrium at least down to $T \simeq 1$ GeV by Higgs portal interactions mediated by the effective operator
\beq
  \mathcal{O}_h \equiv \frac{m_i \tilde{m}_{j}}{m_h^2 f^2} ( \bar{\psi}_i \psi_i) ( \bar{\tilde{\psi}}_j \tilde{\psi}_j) \, ,
  \label{eq:HiggsPortal}
\eeq  
we require $T_{\rm reh} \lesssim \mathcal{O}(\text{GeV})$ and $m_{\rm reh} > M_\Psi$.
The non-thermal yield is  
$Y_\Psi  \sim T_{\rm reh}/m_{\rm reh}  \lesssim 10^{-3}$, giving a similar upper bound as in the thermal production case. The possible contribution to $N_{\rm eff}$ from $\Psi$ decays becomes important and is discussed 
in Appendix~\ref{DR}.

How is the asymmetry actually generated? We are left with various possibilities even given the minimal  particle content (two singlets, one scalar, one twin scalar) depending on the mass spectrum. We assume that only $\Psi_1$ has a significant departure from equilibrium and thus its decay is the main source of the asymmetry, with CP breaking/phases coming either at one loop if $m_\phi< M_1$ (see section \ref{sec:bg1loop}) and at two loops if $m_\phi >  M_1$ (section \ref{sec:bg2loop}). Notice that if $m_\phi > M_2 > M_1$ it is possible for $\Psi_2$ to generate an asymmetry at one loop. It has been shown that a detailed description involving co-annihilations is required to correctly compute the baryon asymmetry \cite{Baldes:2014rda,Baldes:2015lka,Arcadi:2015ffa}, so we leave this case to future work.

\subsection{One-loop model with $M_2> M_1 > m_\phi$} \label{sec:bg1loop}

\begin{figure}
\begin{center}
\includegraphics{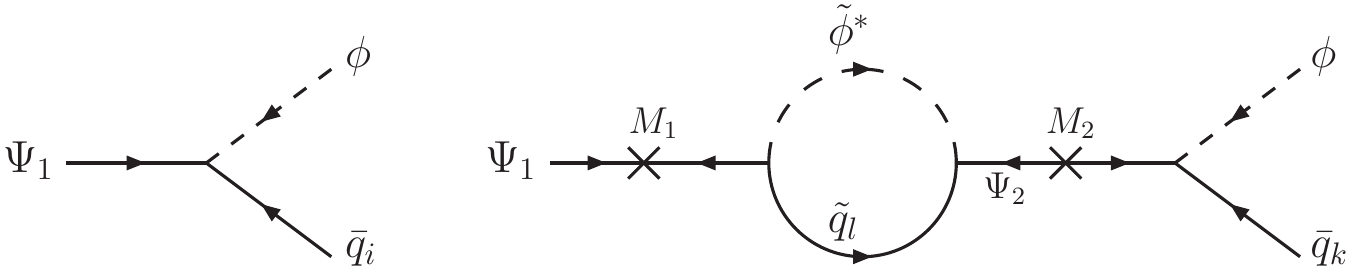}
\caption{Tree-level and 1-loop diagrams for the decays $\Psi_1\to \bar q \phi$ (which will be followed by $\phi\to \bar q\bar q$). Conjugate diagrams contribute to $\bar\Psi_1\to  q \phi^*$. The interference between the diagrams with on-shell particles in the loop gives a non-zero imaginary part. Similar diagrams for $\Psi_1\to \tilde q \tilde\phi^*$ (with twin final states and SM intermediate states) generate the twin asymmetry.
\label{fig:1loop}}
\end{center}
\end{figure}
In this case, since the on-shell two body decays of $\Psi_1$ to $\bar q\phi$ or $ {\tilde q}\tilde\phi^*$ 
are allowed, the baryon asymmetry is generated at one-loop level (see Fig.~\ref{fig:1loop}) as 
 \beq\label{eq:eps1}
 \epsilon = &\epsilon_B =\epsilon_{\tilde B}= \frac{\Gamma(\bar\Psi_1\to q\phi^*) - \Gamma(\Psi_1\to \bar q \phi )}{
 \Gamma(\Psi_1\to \bar q \phi) + \Gamma(\Psi_1\to \tilde q\tilde\phi^*) } \nonumber\\
 \simeq& 
 \left(\frac{M_1}{4\pi M_2}
\right) \left(\frac{\sum_{ij}{\rm Im}(\kappa_{1; i} \kappa_{2; i}^* \tilde\kappa_{1; j} \tilde\kappa_{2; j}^*)}{ \sum_{ij} |\kappa_{1; i}|^2 + |\tilde\kappa_{1;j}|^2}\right). \eeq
The final (twin) baryon abundance is obtained after $\phi^*$ ($\tilde\phi$) decays to $q q$ ($\bar{\tilde q}\bar{\tilde q}$). Note that $\epsilon$ is independent from $\lambda$. As it is from $\kappa_1$ if all the coefficients are of the same order of magnitude: indeed, we obtain
\beq
\epsilon  \sim 10^{-7} \left(\frac{10 \, M_1}{M_2}\right)  \left( \frac{\kappa_{2}}{5 \times 10^{-3}} \right)^2\,\,.
\eeq
Without committing to any $\Psi_1$ production mechanism we can set a lower bound on $\epsilon$ assuming that it either decouples relativistically or is produced non-thermally $(\epsilon>10^{-7})$, and a upper bound  from perturbativity  of the model parameters $(\epsilon<10^{-2})$.
It is therefore clear that $k_2$ cannot be taken too small, $\kappa_2> 5\times 10^{-3}$. On the other hand the decay width of $\Psi_1$ is 
\beq
\Gamma(\Psi_1 \to  \phi \bar q + \tilde{\phi}^* \tilde{q}) = \frac{3}{32 \pi} M_1 \sum_{i} \left(  |\kappa_{1; i} |^2+ |\tilde \kappa_{1; i} |^2 \right)\,\,,
\eeq
and, in the case of thermal production, out of equilibrium decay of $\Psi_1$ dictates
\beq
\kappa_1,\tilde \kappa_1 \lesssim 8 \times 10^{-8} \left(\frac{ M_1}{\tev}\right)^{1/2}   \,\,,
\eeq 
for all couplings of the same order of magnitude. So a hierarchy between $\kappa_1$ and $\kappa_2$ is necessary to generate the baryon asymmetry. If $\Psi_1$  is produced non-thermally at a low reheating temperature,  there is  more freedom and we can take $\kappa_{1}\sim \kappa_{2}$. In any case, constraints on those parameters are coming from the washout effect and requiring decays before BBN,  and more interestingly from dark matter lifetime and collider signatures as will be shown in Sections ~\ref{sec:DMind}-\ref{sec:ColliderPheno}.

\subsection{Two-loop model with $m_\phi > M_1$} \label{sec:bg2loop}
\begin{figure}
\begin{center}
\includegraphics{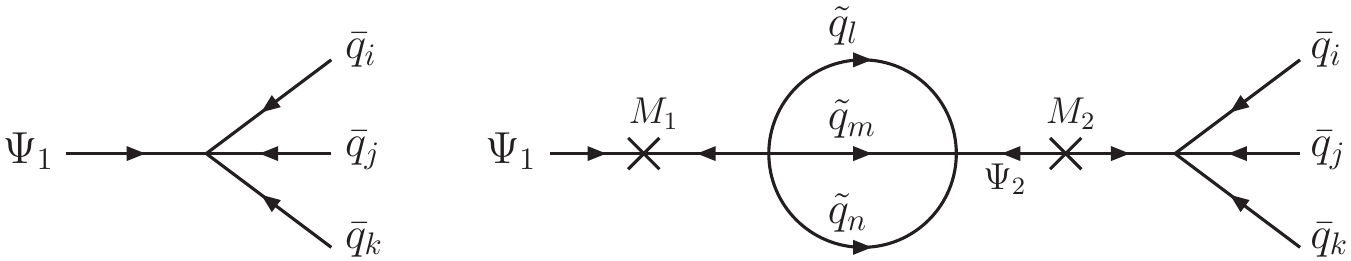}
\caption{Tree level and two-loop diagrams for the decay $\Psi_1\to \bar q\bar q\bar q$. Conjugate diagrams give $\bar \Psi_1\to qqq$. The interference between the two diagrams with on-shell particles in the loop gives a non-zero imaginary part. Similar diagrams for $\Psi_1\to \tilde q\tilde q\tilde q$ generate the twin asymmetry.
\label{fig:2loop}}
\end{center}
\end{figure}

In this case, 
$\Psi_1$ is lighter than $\phi$ so that only three-body on-shell decays to $\bar q\bar q\bar q$ 
 or $\tilde q\tilde q\tilde q$ are allowed. 
To circumvent the Nanopoulos-Weinberg theorem \cite{Nanopoulos:1979gx,RompineveSorbello:2013xwa},
the asymmetry can be generated at two loops  \cite{Monteux:2014hua} as in Fig.~\ref{fig:2loop}.
We get 
\beq\label{eq:eps2}
\epsilon =& \frac{\Gamma(\bar\Psi_1\to qqq) - \Gamma(\Psi_1\to \bar q\bar q\bar q )}{
 \Gamma(\Psi_1\to \bar q\bar q\bar q) + \Gamma(\Psi_1\to \tilde q\tilde q\tilde q) } \nonumber\\
\simeq&\left(\frac{M_1^5}{ 512\pi^3  M_2 m_{\phi}^4}\right)
\left(\frac{\sum_{ijklmn} {\rm Im}(\kappa_{1; i} \kappa_{2;i}^*\tilde\kappa_{2;l}^*\tilde\kappa_{1; l} )|\lambda_{jk}\tilde\lambda_{mn}|^2}{
\sum_{ijklmn}|\kappa_{1;i}\lambda_{jk}|^2 + |\tilde\kappa_{1;l}\tilde\lambda_{mn}|^2}
\right).
\eeq
The second line of Eq.~(\ref{eq:eps2}) is obtained for up-like $\phi$, $\tilde\phi$. 
For the down-like $\phi$, $\tilde\phi$, there could be additional contributions
proportional to  
${\rm Im}(\kappa_{1;i}\kappa_{2;k}^*\tilde\kappa_{2;l}^*\tilde\kappa_{1; n}\lambda_{jk}\lambda_{ji}^*
\tilde\lambda_{mn}\tilde\lambda_{ml}^*)$ from different contractions of fermion operators. 
Barring accidental cancellations all contributions are of the same order.
The width of $\Psi_1$ is now 
\beq
\Gamma(\Psi_1 \to  \bar q\bar q \bar q +  \tilde{q}\tilde{q}\tilde{q}) = \frac{1}{1024 \pi^5} \frac{M_1^5}{m_\phi^4}\ \sum_{ijk} \left(  |\kappa_{1; i} \lambda_{jk} |^2+ |\tilde \kappa_{1; i} \tilde \lambda_{jk} |^2\right)\,\,,
\eeq
and so $\kappa_1 \lambda \lesssim 10^{-5}$  for out-of-equilibrium decay. Again, $\kappa_1$ must be small in the thermal production scenario, and can be ${\cal O}(1)$ in the non-thermal case.
In contrast with the one-loop model,  $\epsilon \propto {\cal O}(\kappa_2^2\lambda^2 )$, so that relatively large $\lambda$'s are required. Moreover because of the small numerical prefactor in $\epsilon$, successful baryogenesis also requires $M_1 \gtrsim 0.3 \{m_{\phi},M_2\}$. Such small mass splitting and large couplings are the generic consequence of this type of baryogenesis. We will investigate phenomenological implications in Section \ref{sec:ColliderPheno}.


\section{Unstable DM, Indirect and Direct Detection}
\label{sec:DMind}

In the present framework, the $U(1)_{B-\tilde B}$ symmetry predicts equal asymmetries in our sector and in the twin sector. The lightest twin quarks hadronize into twin baryons which form the dark matter. Recalling that only $U(1)_{B-\tilde B}$ is conserved, a 5\gev\ dark matter is unstable and decays into SM hadrons. In the Lagrangian, integrating out $\phi,\tilde\phi$ and $\Psi_A$ results in the $B$,$\tilde B$ violating operator
\beq\label{eq:DMlagr}
{\cal L}_{B-\tilde B}=& 
\frac{ \kappa_{A; i}\tilde \kappa_{A;l}\lambda_{jk}\tilde\lambda_{m n}}{  M_A m_\phi^2m_{\tilde \phi}^2}
(q_i q_j q_k)(\tilq_{l} \tilq_{m} \tilq_{n}) +h.c.\,.
\eeq
While other options are interesting and worthwhile studying, in the following we will assume  the simplest, most typical scenario in which that the dark matter is composed of twin neutrons, $\tilde B_{DM}=\tilde n=(\tilde u\tilde d\tilde d)$. 
Then, at a scale well below the twin EW scale,
the relevant quarks for the twin neutron decay would be $\tilde u$, $\tilde d$, $\tilde s$, 
whose masses are smaller than $\tilde \Lambda_{QCD}$.
The effective operator after integrating out high momentum contributions can be written as
\beq\label{eq:qqqO}
{\cal L}_{decay} =\sum_{  \tilde{\cal O}}\frac{\kappa_{A;i}\tilde\kappa_{A;l}\lambda_{jk}\lambda_{lm} {\cal F}_{\cal O}^{lmn} }{M_A m_\phi^2 m_{\tilde\phi}^2} 
 (q_j q_k) ( q_i{ \tilde{\cal O}}),
\eeq
where $\tilde {\cal O} = \tilde d^c (\tilde u^c \tilde d^c),\, \tilde s^c(\tilde u^c \tilde d^c),\, 
\tilde u^c(\tilde d^c\tilde s^c)$. 
Possible flavor violation  induced by short-distance physics (exchange of $\tilde W$) 
is included in a new coefficient $\FO$.  Flavor violating effect become important 
when  $\tilde\kappa_A$, and $\tilde\lambda$ are too small for the first and second generation. As an example, 
for an up-type scalar, because of $\tilde\lambda_{ii}=0$,  one could not have the operator $\tilde  u(\tilde d\tilde d)$. Instead, we can have $\tilde t(\tilde b\tilde d)$, $\tilde c(\tilde b \tilde d),\ldots$ which turn into the operator $\tilde d(\tilde u\tilde d)$ by twin $W$ exchange at one-loop level.  The magnitude of flavor violating ${\cal F}_{d(ud)}$ will depend on the twin masses and twin CKM. Assuming $Z_2$-symmetric Yukawas (which gives $\widetilde V_{CKM}=V_{CKM}$), 
the largest contribution to ${\cal F}_{ d(  u  d)}$ comes from $\tilde u_{l}\tilde d_{m}\tilde d_n=\tilde c  \tilde b\tilde d$ and is 
\beq
&{\cal F}_{ d( u d)}^{ c b d} \sim \frac{\tilde \alpha_2}{4\pi}\frac{ m_{\tilde c}m_{\tilde b}}{ m_{\tilde W}^2} \log{\frac{m_{\tilde c}^2}{ m_{\tilde W}^2}}(\widetilde V_{CKM})_{\tilde c\tilde d} (\widetilde V_{CKM})_{\tilde u\tilde b}\sim 10^{-7}.
\eeq
For loops involving the twin top,\footnote{
The full expression  is ${\cal F}_{ d( u d)}^{lm n }=\delta^{lmn}_{ u  d d}+\frac{\tilde \alpha_2}{4\pi} (\widetilde V_{CKM})_{l \tilde d} (\widetilde V_{CKM})_{\tilde u{m }}\, m_{\tilde u_{l }} m_{\tilde d_{m }}F_W(m_{\tilde u_{l }},m_{\tilde d_{m }})+(m \leftrightarrow n )$, where $F_W$ is defined as $ F_W(m_u,m_d)\equiv
 \frac1{(m_d^2-m_u^2)}\left(\frac{m_d^2}{m_d^2+M_W^2}\log\frac{m_d^2}{M_W^2}-\frac{m_u^2}{m_u^2+M_W^2}\log\frac{m_u^2}{M_W^2}\right)$ for heavy twin quarks and one should replace $m_{\tilde u,\tilde d,\tilde s}$ with $ \tilde\Lambda_{QCD}$ when the twin quarks in the loop are light. 
 For a down-type $\tilde\phi$, the difference is just to change $lmn\to mln$. The other operators $\FO$ have similar expressions, with different indices for $\widetilde V_{CKM}$.}
 the largest channel has $\tilde u_{l}\tilde d_{m}=\tilde t\tilde b$ and yields ${\cal F}_{d(ud)}^{ t b d}\sim 10^{-8}$.  
Without committing ourselves to a specific twin flavor structure, we will keep $\FO$ as a free parameter and keep the above order-of-magnitude estimate in mind.

In Appendix~\ref{app:chiral} we present a detailed derivation of the low energy effective Lagrangian resulting from Eq.~(\ref{eq:qqqO}). 
The interactions relevant for DM decay, up to first order in $1/f_{\tilde\pi}$, can be schematically written as
\beq\label{eq:DMchiralsimple}
{\cal L}_{\tilde n}=\tilde\Lambda_{QCD}^3 c_{\tilde n} (\tilde n q)(qq) 
+ i \tilde\Lambda_{QCD}^3 \frac{c_l\tilde M_l}{f_{\tilde\pi}} (\tilde n  q)(qq),
\eeq
where $f_{\tilde\pi}$ is the twin pion decay constant, with $f_{\tilde\pi}=f_\pi \tilde \Lambda_{QCD}/\Lambda_{QCD}$ and $f_\pi=130$ MeV,
the twin confinement scale is fixed by $\widetilde \Lambda_{QCD}= (m_{\tilde B}/m_{p})\Lambda_{QCD} \simeq1.35\gev$, and $\tilde M_l=\tilde M_l=\tilde \pi^0,\, \tilde\eta^0,\, \tilde K^0$ are twin light mesons; the $c$ coefficients have dimension of mass$^-5$ and  are related to the ones of Eq.~(\ref{eq:qqqO}) by ${\cal O}(1)$ coefficients arising from the hadronic matrix elements. We refer the reader to appendix~\ref{app:chiral} for the explicit interaction Lagrangian, Eq.~(\ref{eq:DMchiral}), 
and the relevant matrix elements, Eq.~(\ref{eq:twinME}).
 
From the interaction in Eq.~\eqref{eq:DMchiralsimple}, different decay channels are possible:
\begin{itemize}
\item $\tilde B_{DM} \to q_i q_j q_k$: decay to SM quarks lighter than the dark matter, that is, $u,d,s,c$. Decays with $b$ quarks are suppressed by the presence of off-shell $B$ mesons or baryons, all heavier than the dark matter.
Each final quark energy is around or above 1\gev, so the parton level decay rate  gives a good approximation at leading order. The decay rate is
\beq\label{eq:DMgamma}
\Gamma_{DM\to qqq}=\frac{\mdm^5}{1024\pi^3} \left( \frac{\kappa_{A; i}\tilde \kappa_{A;l } \lambda_{jk}\tilde\lambda_{m  n } {\cal F}_{d(ud)}^{lm  n } \widetilde \Lambda_{QCD}^3 }{M_A m_\phi^4 } \right)^2\,,
\eeq
\item $\tilde B_{DM} \to q_i q_j q_k \tilde M_l$. The decay rate to three light quarks and a twin meson will be 
\beq\label{eq:DMgamma2}
\Gamma_{DM\to qqq\tilde M}\simeq  \frac1{16\pi^2}\frud{m_{DM}}{f_{\tilde\pi}}^2
\left(\frac{\mdm^5}{1024\pi^3}\right) \left( \frac{\kappa_{A; i}\tilde \kappa_{A;l } \lambda_{jk}\tilde\lambda_{m  n } {\cal F}_{\cal O}^{lm  n } \widetilde \Lambda_{QCD}^3 }{M_A m_\phi^4 } \right)^2\,\,.
\eeq
Compared to $\Gamma_{DM\to qqq}$, $\Gamma_{DM\to qqq\tilde M}$ has an enhancement due to $m_{DM}/ f_{\tilde\pi} \sim 7$, and a phase space suppression (due to four-body vs three-body decays, and also to large final state masses). The two rates are about the same order of magnitude for similar value of $\FO$. It is possible  that decays to three quarks are flavor-suppressed while decays to twin mesons are not (e.g. for up-type $\phi$, the three-quark decay is suppressed by ${\cal F}_{ d (  u  d)}\sim 10^{-7}$ while the decay with one twin kaon is not, ${\cal F}_{ u ( d  s)}=1$).\end{itemize}

The lifetime obtained from Eqs.~\eqref{eq:DMgamma}, \eqref{eq:DMgamma2} can be written as 
\beq
\tau_{DM} \sim 10^{18}\sec \frud{m_\phi}{1\tev}^8\frud{5\gev}{m_{DM}}^{5}\frud{M_A}{1\tev}^2
\left(\frac{(10^{-1})^2}{\kappa_{A; i}\tilde \kappa_{A;l }}\, \frac{(10^{-2})^2}{\lambda_{jk}\tilde\lambda_{m n }}\right)^2   
\frud{1}{{\cal F}_{\cal O}^{lmn}}^2. \,\,
\label{eq:DMtau}
\eeq
where it is understood that the lifetime is reduced when multiple decay channels are accessible. This estimate applies to decays to light quarks ($u,d,c,s$); on the other hand, $m_{DM}$ is barely above $m_b$, and decays involving $b$ quarks are heavily suppressed: because the $b$ quark has an appreciable lifetime, it will hadronize and one should consider the lightest $b$ hadrons $B,B_s,\Lambda_{b}$, which have masses of $5.28,\ 5.36$ and $5.6$\gev\ respectively. Thus, a 5\gev \ dark matter decaying to $b$'s would entail off-shell $b$ hadrons, suppressed by the $b$ decay rate, which contains CKM factors and $M_W$: for flavor universal couplings, we estimate
\beq
\frac{\Gamma_{DM\to \bar p \pi^+}}{\Gamma_{DM\to \bar p B^+}}\sim\frac{1}{8\pi}\frac{\Gamma_B}{m_B}\simeq10^{-14}\,.
\eeq
Therefore, decays to light quarks dominate unless a huge hierarchy appears in the couplings, e.g. $\lambda_{bs}/\lambda_{ds}\gg 10^7$, which seems highly unnatural if these new couplings originate from the same source as known quark masses and mixings.\footnote{For example, in an MFV setup, the largest hierarchy achievable would be $\lambda_{bs}/\lambda_{ds}\sim m_b/m_d \sim 10^3 \ll 10^7$.}

The lifetime computed from Eq.~(\ref{eq:DMtau}) is easily larger than the age of the universe, $\tau_0=4.35\times 10^{17}\sec$, so it seems that the (unstable) twin baryon can form all of the dark matter at present times. 
On the other hand this is just a necessary but not sufficient condition, as dark matter decaying to hadronic final states will produce a host of indirect detection signals, including gamma rays, antiprotons, antideuterons and positrons, yielding much stronger limits on the lifetime. In the following, we use limits on gamma rays obtained from the Fermi-LAT satellite \cite{FERMILAT,Bregeon:2013qba} as they are independent from astrophysical uncertainties affecting the intergalactic propagation of charged particles. In particular, we adopt from Ref.~\cite{Massari:2015xea} the bounds on the decay mode DM$\to u \bar u$ to constrain the decay $\tilde B_{DM}\to \bar q_i\bar q_j\bar q_k$, where the final state can involve all first- and second-generation quarks, $u,d,c,s$, as the gamma ray spectra are similar for both decays. At $m_{DM}=5\gev$ we take the lower limit on the DM lifetime 
\beq
\tau_{DM}^{obs.}> 10^{25}\sec,
\eeq
which can be compared to the twin baryon lifetime from Eq.~\eqref{eq:DMtau}. It is immediately seen that small couplings are necessary to satisfy the constraints on decaying dark matter. 

In contrast to upper bounds from DM stability, requiring a sizable baryon asymmetry gives lower bounds. Notice that while dark matter decays only involve couplings to the first two generations, the asymmetry parameter depends on all generations. Thus, dark matter stability and a successful asymmetry generation are tightly related to each other only when additional assumptions on the flavour structure are made, e.g. flavour universal versus hierarchical couplings. Each case gives a distinct set of collider signatures that can be effectively probed at the LHC. We will study the interplay between baryogenesis, dark matter lifetime and LHC signatures in the next section. 

Finally a comment is in order about direct detection. Apart from Higgs portal interactions, which are suppressed by the Yukawas, we must take into account operators of the form $\tilde q \bar{\tilde q }q \bar q$ obtained by one-loop box diagrams involving the exchange of $\phi$ and $\Psi$. Their magnitude can be estimated as $\sigma_{SI} \sim  (10^{-39}- 10^{-41}) (\kappa_1^2{\rm TeV}/M_1)^4  {\rm cm}^{2}$, and
so it could be within the sensitivity of future direct detection experiments \cite{Akerib:2015cja}. 
This kind of diagram also contributes to the kinetic mixing between $U(1)_Y$ and the twin hypercharge (if gauged) at three-loop level, but its effect is safe from the present constraints on milli-charged particles \cite{Davidson:2000hf}.
Finally our scenario also predicts induced nucleon decays, with proton lifetime estimated at around $10^{32}$ years \cite{Davoudiasl:2010am,Davoudiasl:2011fj}. The process $\tilde n B \to M \tilde M$ can also induce additional signal at direct detection experiments \cite{Huang:2013xfa} but depends on the physics associated with the twin meson. We leave a detailed study of all the possible direct and indirect detection phenomenology to future work.


\section{Collider phenomenology}
\label{sec:ColliderPheno}
We are now ready to study the consequences at LHC.
As discussed in the previous section, constraints on the dark matter lifetime force the presence of small couplings in the theory ($\tau_{DM} \propto (\kappa \lambda)^{-4}$), while baryogenesis gives lower bounds. If the Standard Model flavor structure is to be explained by a symmetry, it is likely that the new couplings $\kappa, \lambda$ also inherit a flavor hierarchy from the same symmetry. In a $Z_2$-symmetric model, the same structure will apply to the twin sector. With this in mind and for the sake of simplicity, we will assume that the $\lambda$ and $\kappa$ couplings preserve the $Z_2$ symmetry between our sector and the twin sector, $\lambda=\tilde \lambda, \kappa=\tilde\kappa$.
We will study two minimal benchmark scenarios:
\begin{itemize}
\item For $m_\phi<M_1$, the one-loop asymmetry is proportional to $\kappa_2^2$ only: to satisfy the DM lifetime constraint, we can take the $\lambda$ couplings to be small, which in turns can result in a sizable decay length of $\phi$. For simplicity, we assume no flavor structure with couplings of similar magnitude for different flavor indices, $\kappa_{A;i}\sim \kappa_{A},\ \lambda_{jk}\sim \lambda$.
\item For $m_\phi> M_1$, the two-loop asymmetry is proportional to $\kappa_2^2\lambda^2$. Having a stable DM candidate compels a flavor hierarchy on the couplings, with those involved in baryogenesis required to be at least $\mathcal O(0.1)$, or larger, while the twin baryon decay rate is suppressed by smaller couplings  involving first and second generations, and for example $\lambda_{jk}\tilde\lambda_{m_0n_0}<(10^{-4})^2$ can be sufficient to raise the DM lifetime. 
As benchmark points for a possible origin of the (twin) flavor structure, we will take Minimal Flavor Violation (MFV) \cite{D'Ambrosio:2002ex}  or a horizontal Abelian symmetry {\it \`a la} Froggatt-Nielsen \cite{Froggatt:1978nt}.
We will not commit to a particular flavor model, but use those as typical examples.
\end{itemize}
In the rest of this section, we will analyze the experimental signatures and use public LHC searches to set limits on  these scenarios. 
At the LHC, the colored scalar $\phi$ can be pair-produced via gluon fusion, $gg\to \phi\phi^*$, as well as resonantly produced from quarks, $q_i q_j \to \phi^*$ (if $\lambda_{ij}$ is large), while direct production of the Dirac singlet $\Psi_1$ is negligible. The cross sections are the same as for a top squark (stop) in SUSY, while the $\phi qq$ couplings mimics baryonic $R$-parity violation (RPV). The main experimental signatures are single and paired di-quark resonances from direct $\phi\to \bar q\bar q$ decays, as well as four-quark resonances from cascade decays involving the Dirac singlet, $\phi\to q\Psi_1\to q(\bar q\bar q\bar q)$ (only two-loop case). Depending on the flavor structure of the operators, some of the final states might be top quarks. Finally, when the singlet decays into twin quarks, $\Psi_1\to \tilde q\tilde q\tilde q$, the twin quarks leave the detector as missing energy: resonant production gives a monotop (or monojet) while pair-production gives $q\bar q+\slashed{E}_T$ (again, similar to SUSY signatures).

The signatures are similar to those of a right-handed stop with a bino-like (unstable) neutralino with baryonic RPV couplings $\lambda_{3jk}''$, except for (i) decays into the twin sector resulting in missing energy and (ii) the absence of same-sign top signatures which arise from the Majorana gaugino mass. The RPV scenario was recently studied by one of the authors in Ref.~\cite{Monteux:2016gag}: here, we refer to that study for details about interpreting the experimental searches and include  baryogenesis and DM lifetime to further constrain the present model. 

If the decays are prompt, the relevant searches are the CMS {\it scouting} resonant dijet search \cite{CMS:2015neg} and the ATLAS Gaussian dijet search \cite{Aad:2014aqa}, together with searches for paired dijet resonances \cite{Khachatryan:2014lpa,ATLASCONF2015026}, which exclude $\phi \to qq$  between 200\gev \ and 350\gev, and, when combined, $\phi\to b q$ between 100\gev \ and 385\gev. For pair-produced scalars decaying via $\Psi_1$ to $\bar t tjj$, we use the ATLAS four-top search \cite{Aad:2015kqa}. These searches are all based on 20 fb$^{-1}$ of 8\tev\ collisions.

Likewise, there are strong experimental limits on long-lived particles decaying to jets: we use recasted analyses \cite{Liu:2015bma,Csaki:2015uza} based on 8\tev\ searches \cite{CMS:2014wda,Aad:2015rba} for displaced $\phi$ decays and the recent 13\tev\ search \cite{CMS:2015kdx} for collider-stable $\phi$'s. In particular, we quote the displaced stop limits in \cite{Liu:2015bma}. We emphasize that the long-lived dijet signatures are common to both scenarios of up-type and down-type $\phi$, as even a top final state would result in a $b$-jet, and the searches do not require the displaced jets to form a resonance.

\subsection{One-loop model with flavor universal couplings} 
In this section we study the one loop flavour-universal benchmark. We rewrite the asymmetry parameter in Eq.~\eqref{eq:eps1} by summing over flavor indices and explicitly defining a CP phase $\varphi_{CP}$,
\beq\label{epsB1-loop}
\epsilon\simeq 1.2 \times 10^{-2} \kappa_{2}^2  \,\sin\varphi_{CP} \frud{10M_1}{M_2}\,.
\eeq
The LHC lifetime of the (up-type) $\phi$ scalar is
\beq\label{eq:phictau}
c\tau(\phi\to \bar q \bar q)= 1.7 \text{ mm} \frud{10^{-7}}{\lambda}^2\frud{100\gev}{m_\phi}\,,
\eeq
where we have summed over $ij=ds,db,sb$; a down-type scalar can decay into more final states, including a top, so the lifetime will be suppressed by a factor of order two.
Combining Eqs. \eqref{epsB1-loop} and \eqref{eq:phictau} with the DM lifetime, Eq.~\eqref{eq:DMtau}, we find 
\beq\label{eq:taudm1-loop}
&\tau_{DM}=2.7 \times 10^{26}\text{sec} \
\frac{\sin^2\varphi_{CP}}{{\cal F}^2}
\frud{M_1}{1\tev}^2\frud{m_\phi}{0.1\tev}^{10}\frud{10M_A/\kappa_A^2}{M_2/\kappa_2^2}^2
 \frud{c\tau}{1\text{mm}}^2\frud{10^{-2}}{\epsilon}^2\,.
\eeq
Here we didn't specify the flavor index for ${\cal F}$. 
If the asymmetries are generated by the decay of thermally produced $\Psi_1$, 
$M_A/\kappa_A^2$ is dominated by $M_2/\kappa_2^2$. 
Otherwise, if $\Psi_1$ is produced non-thermally and assuming $\kappa_1\sim \kappa_2$, then $M_A/\kappa_A^2$ is dominated by $M_1/\kappa_1^2$. 
We have also set $m_{DM}=5\gev$. The absence of a flavor hierarchy ensures  that DM decays to light quarks dominate over decays via off-shell $b$-hadrons. At the LHC, the $\phi$ decays will be democratic, including to bottom and top quarks (if kinematically allowed).

As the dark matter lifetime heavily depends on the magnitude of the twin flavor violation parameter ${\cal F}$, we consider separately the cases in which $\phi$ is up-type ($Q_\phi=2/3$) or down-type ($Q_\phi=-1/3$):
\begin{itemize}
\item up-type $\phi$: in this case, the three-quark dark matter decay is suppressed by a factor of $({\cal F}^{ c b d}_{d( u d)})^2\sim10^{-14}$, while the decay with a twin kaon needs no flavor violation, ${\cal F}^{uds}_{ u( d  s)}=1$ and therefore dominates. 

\item down-type $\phi$: in this case the three-quark decay rate is unsuppressed, ${\cal F}^{dud}_{\tilde d(\tilde u d)}= 1$. 

\end{itemize}
In both cases, from Eq.~\eqref{eq:taudm1-loop} and given $\tau_{DM}^{obs.}\gtrsim 10^{25}\sec$, either displaced vertices, heavier mass scales or small $\epsilon$ are required. The dependence on $\epsilon$ can be translated into a dependence on the primordial $\Psi_1$ abundance via the observed matter-antimatter asymmetry, $ Y_{\Delta B}=\epsilon Y_{\Psi_1} \simeq 10^{-10}$. A small asymmetry parameter corresponds to a large $\Psi_1$ abundance, with $Y_{\Psi_1}\lesssim 10^{-3}$ as discussed in Section \ref{thermalH}.

Given universal flavor-violating interactions, one might worry about FCNCs: the phenomenology is again similar to that of RPV SUSY \cite{Barbier:2004ez}, except for the automatic absence of $\Delta B=1$ processes such as $n-\bar n$ oscillation and dinucleon decay. $K-\bar K$ oscillation gives the strongest constraint, with $|\lambda|^2<6\times 10^{-4}\frud{m_\phi}{100\gev}$, that is, $\lambda\lesssim 10^{-2}$. On the other hand, the dark matter lifetime \eqref{eq:taudm1-loop} implies $\lambda\lesssim 10^{-7}$, so that FCNCs are always negligible.

At the LHC, the limits  from paired dijet resonances are slightly weaker than in the original searches, as $\phi$ decays democratically to all quarks, thus reducing the efficiency of the searches requiring $b$-tagging. The CMS results for $\phi \to qq$ still apply, while the smaller branching ratio lowers the $b$-tagged limits to $m_\phi\gtrsim 350\gev$. Paired dijet resonances appear for both up- and down-type $\phi$ (for the latter,  $\phi\to tj$ decays are negligible at low masses where the limits apply).

\begin{figure}[t]
\begin{center}
\includegraphics[width=0.9\textwidth]{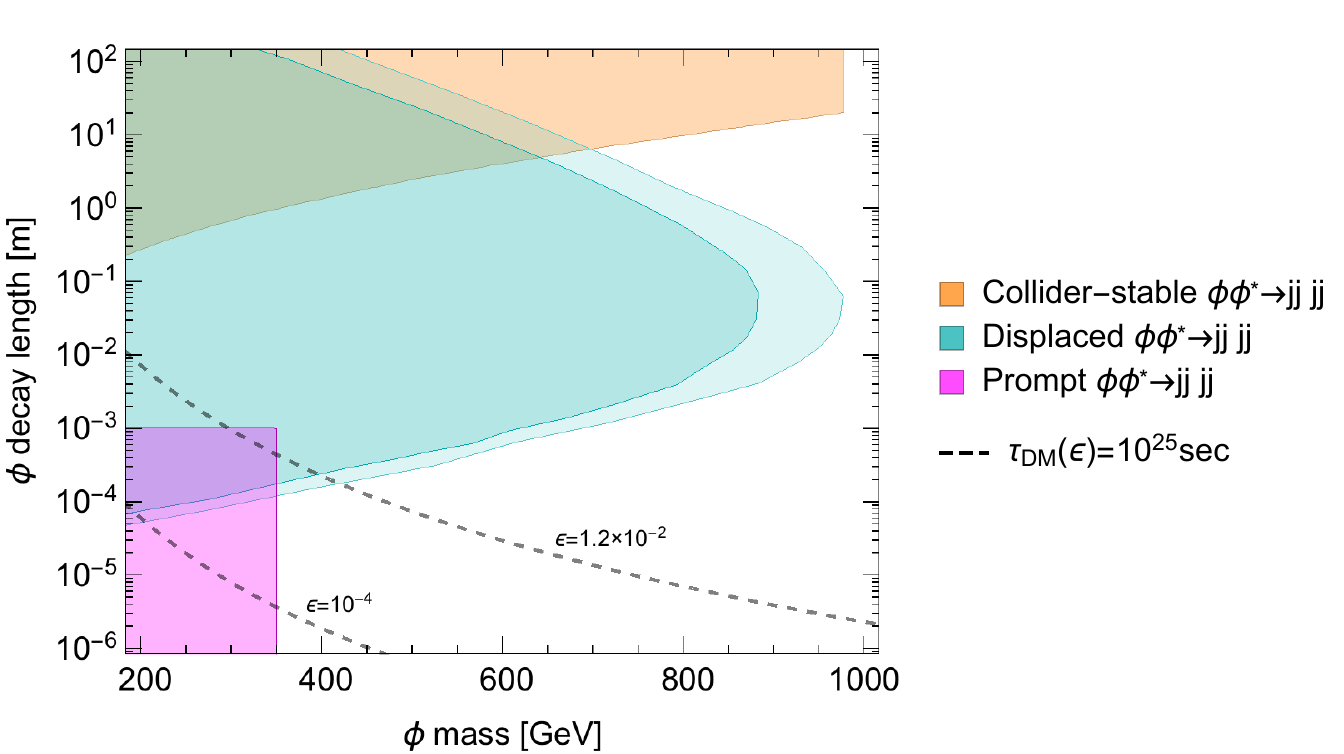}
\caption{LHC and dark matter lifetime constraints on the one-loop model for baryogenesis with universal couplings and having set $M_1=1\tev, M_2=10M_1$. Shaded regions are excluded by the different experimental signatures. LHC constraints on a pair-produced scalar $\phi$ decaying into jets are respectively shown in magenta, blue and orange for prompt \cite{Khachatryan:2014lpa,ATLASCONF2015026}, displaced \cite{Liu:2015bma} and collider-stable \cite{CMS:2015kdx} decays. The displaced exclusions are adapted from Ref.~\cite{Liu:2015bma}, and agrees well with Ref.~\cite{Csaki:2015uza}; the lighter blue area corresponds to changing the acceptances up/down by 1.5.
Dashed lines saturate the DM lifetime bound $\tau_{DM}\geq10^{25}\sec$: for each value of the asymmetry parameter $\epsilon$, regions below the dashed line are excluded. The DM lifetime constraints are shown assuming decays predominantly mediated by $\Psi_1$ and $\kappa_1=\kappa_2$ (corresponding to non-thermal $\Psi_1$): for the thermal case the constraints are slightly weaker.
}
\label{fig:1loopctau}
\end{center}
\end{figure}

We show the resulting constraints in Figure \ref{fig:1loopctau}, where for reference we have set $M_1=~1\tev$, $M_2=10M_1$ and shaded regions are excluded by different experimental signatures. 
From the top of the figure downward, in orange, cyan and purple we respectively show regions excluded by searches for heavy stable charged particles at 13\tev\ \cite{CMS:2015kdx}, displaced dijets at 8\tev\ \cite{CMS:2014wda,Aad:2015rba} as recasted in \cite{Liu:2015bma} (the light shaded area reflects an $O(1)$ variation on experimental acceptances) and prompt paired dijets at 8\tev \ \cite{Khachatryan:2014lpa,ATLASCONF2015026}.
Dashed lines saturate the DM lifetime constraints from Eq.~\eqref{eq:taudm1-loop}, for increasingly small values of the asymmetry parameter: $\epsilon=1.2\times 10^{-2}$ corresponds to taking all couplings and phases to be O(1) in Eq. \eqref{epsB1-loop}, while there are no limits for $\epsilon\simeq 10^{-7}$, corresponding to the upper limit on the primordial $\Psi_1$ abundance $Y_{\Psi_1}\simeq10^{-3}$. We took $\kappa_1=\kappa_2$ so that DM decays are mediated by $\Psi_1$: this corresponds to baryogenesis from non-thermally produced  $\Psi_1$'s. In the thermal case, $\kappa_1\ll1$ and DM decays are dominated by $\Psi_2$, giving slightly weaker limits on $\tau_{DM}$.

Sub-TeV masses for $\phi$ are mostly excluded by a combination of LHC and decaying DM searches, unless the parameters setting the baryon asymmetry are also small, e.g. $\kappa< 10^{-2}$, for which prompt decays are still allowed at masses above 400\gev. It will be interesting to see how 13\tev \ searches for paired dijet resonances will cover this region. 

\subsection{Two-loop model with flavor hierarchy}
We now turn our attention to the case $M_1<m_\phi$, where the baryon asymmetry is generated at two loops and a hierarchy in couplings is required to stabilize the dark matter while keeping $\epsilon$ large enough. In particular, couplings involving first-and second-generation couplings (contributing to DM decays) need to be suppressed with respect to couplings involving $t,b$ (heavier than the DM).

We rewrite the asymmetry in Eq.~\eqref{eq:eps1} by also defining a phase $\varphi_{CP}$,
\beq \label{epsB2-loop}
\epsilon
\simeq 3.2\times 10^{-6}\kappa^2_{2;I} \lambda_{JK}^2 \sin\varphi_{CP}\frud{M_1}{m_\phi}^4\frud{10M_1}{M_{2}}\,,
\eeq
where $(IJK)$ are the flavor indices relevant for baryogenesis (the possibilities being $q_I q_J q_K=tbs, cbs, btb$ or $bcb$) with 
\beq
M_1> \max\Big[\, m_{q_I}+m_{q_J}+m_{q_K}, \, m_{\tilde q_I}+m_{\tilde q_J}+m_{\tilde q_K}\,\Big] 
\eeq for all of the $\Psi_1$ decay products to be on-shell. 
The twin quarks have $m_{\tilde q}=(f/v) m_q$, so we expect $M_1\gtrsim 500\gev$ if the top and twin top take part in baryogenesis, while $\Psi_1$ could be lighter otherwise (although it would be hard to justify an O(1) coupling involving the charm while keeping the coupling for the top perturbative).
Given the upper bound on the $\Psi_1$ abundance, $Y_{\Psi_1}\lesssim 10^{-3}$, we require  $\kappa_{2;I}^2\lambda_{JK}^2 \gtrsim 0.1$, $M_1\gtrsim 0.5 m_\phi$ and $\sin\varphi_{CP}\gtrsim 10^{-1}$. 

Because of the large twin top mass, there will be some phase space suppression in the computation of $\epsilon$ if $M_1$ is below a TeV. This suppression can easily be $10^{-1}$ or less for $M_1<2m_{\tilde t}$, in which case even O(1) couplings cannot generate enough baryon asymmetry. Hence, a heavier singlet will be preferred.

As before, we can tie together the DM lifetime and the asymmetry parameter:
\beq\label{eq:tauDM2loop}
&\tau_{DM}= 1.65\times 10^{25}\text{sec}\ \sin^2\varphi_{CP} \frud{M_1}{500\gev}^{10}\frud{10M_A}{M_2}^2 \frud{10^{-7}}{\epsilon}^2 \frud{10^{-7}}{{\cal F}}^2 \left(\frac{10^{-3}}{\mathcal R}\right)^2,
\eeq
where we defined a flavor ratio, $\mathcal R$, that takes into account the different couplings entering in DM decays (numerator) and in baryogenesis  (denominator):
\beq
\mathcal R \equiv \frac{\kappa_{A;i}\tilde \kappa_{A;l_0}}{\kappa_{2;I}\tilde\kappa_{2;I}}\frac{\lambda_{jk}\tilde\lambda_{m_0n_0}}{\lambda_{JK}\tilde\lambda_{JK}}\,.
\eeq
Here we indicated by $(l_0m_0n_0)$ the indices of the twin operator most relevant for DM decay, by $(IJK)$ those relevant for baryogenesis and by $(ijk)$ the indices of the dominant SM final states in the decay. In our flavor examples (MFV and horizontal symmetries, with $Z_2$-symmetric twin Yukawas), they are 
\begin{itemize}
\item  for an up-type scalar, $q_Iq_Jq_K=tbs, q_iq_jq_k=cds$. For dark matter decays to three quarks, $\tilq_{l_0}\tilq_{m_0}\tilq_{n_0}=\tilde t\tilde b \tilde s$, ${\cal F}\sim10^{-8}$ and $\cal R$ is $10^{-5}$  assuming MFV, or $10^{-2.5}$ with a horizontal symmetry. For dark matter decays with a twin kaon, $\tilq_{l_0}\tilq_{m_0}\tilq_{n_0}=\tilde u\tilde d \tilde s$, ${\cal F}=1$ and $\cal R$ is $10^{-13}$  or $10^{-7.5}$ for the two flavor {\it ansatze}.
\item for a down-type $\phi$, $q_Iq_Jq_K=btb,q_iq_jq_k=scs$. 
For decays to three quarks, we can have the decay without flavor violation, ${\cal F}=1$ and $\tilq_{l_0}\tilq_{m_0}\tilq_{n_0}=\tilde d\tilde u \tilde d$, which gives ${\cal R}=10^{-16}$ assuming MFV and  ${\cal R}=10^{-9}$ with a horizontal symmetry, or a flavor-violating loop factor ${\cal F}\sim10^{-8}$ with $\tilq_{l_0}\tilq_{m_0}\tilq_{n_0}=\tilde b\tilde t \tilde b,$ yielding ${\cal R}_{MFV}=10^{-3.3}$ and  ${\cal R}_{horiz.}=10^{-1.3}$. For decays with a twin kaon, we have $\tilq_{l_0}\tilq_{m_0}\tilq_{n_0}=\tilde d\tilde u \tilde s$, ${\cal F}=1$ and ${\cal R}_{MFV}=10^{-15}$ and ${\cal R}_{horiz.}=10^{-8}$.
\end{itemize}
The potentially small factors  ${\cal F}$ and $\mathcal R$ make dark matter lifetime rather dependent on the flavor dynamics: without committing to a specific flavor model, we will keep ${\cal F}\mathcal R$ as an independent variable, and note that DM lifetime constrains ${\cal F}\mathcal R\lesssim 10^{-10}$, while keeping in mind that typical values for that quantity are between $10^{-8}$ and $10^{-12}$. Therefore, reasonable flavor assumptions give values of the DM lifetime at around the experimental upper bounds.

At the LHC, in addition to QCD production via gluon fusion (again resulting in a RPV stop-like phenomenology), the resonant production channel $qq\to \phi$ via a large coupling $\lambda_{JK}$ is also possible. Apart from dijet resonances, the most interesting new signature is a monotop:
\beq
q_Jq_K\to \phi^* \to \bar t\bar \Psi_1\to \bar t + \slashed{E}_T
\eeq
which arises because the $\Psi_1$ also decays to twin quarks: $
{\rm Br}(\Psi_1\to\tilde q\tilde q\tilde q)\sim {\rm Br}(\Psi_1\to \bar q\bar q\bar q)
$, with some kinematic suppression if the top partner mass is near that of $\Psi_1$.
There are additional signatures such as $gg \to t\bar t+4j$, or $gg\to tt\bar t+2j$, where it could be possible to overcome the large SM $t\bar t$ background by requiring large $b$-jet multiplicities. We will not focus on those as they are not covered by existing searches. Because large couplings are required from baryogenesis, all decays are prompt, except for the thermal case where out-of-equilibrium $\Psi_1$ decays result in displaced vertices and are excluded below approximately 1\tev~\cite{Cui:2014twa,Csaki:2015uza}. 

\begin{figure}[t]
\begin{center}
\makebox[\textwidth][c]{\includegraphics[height=0.38\textwidth]{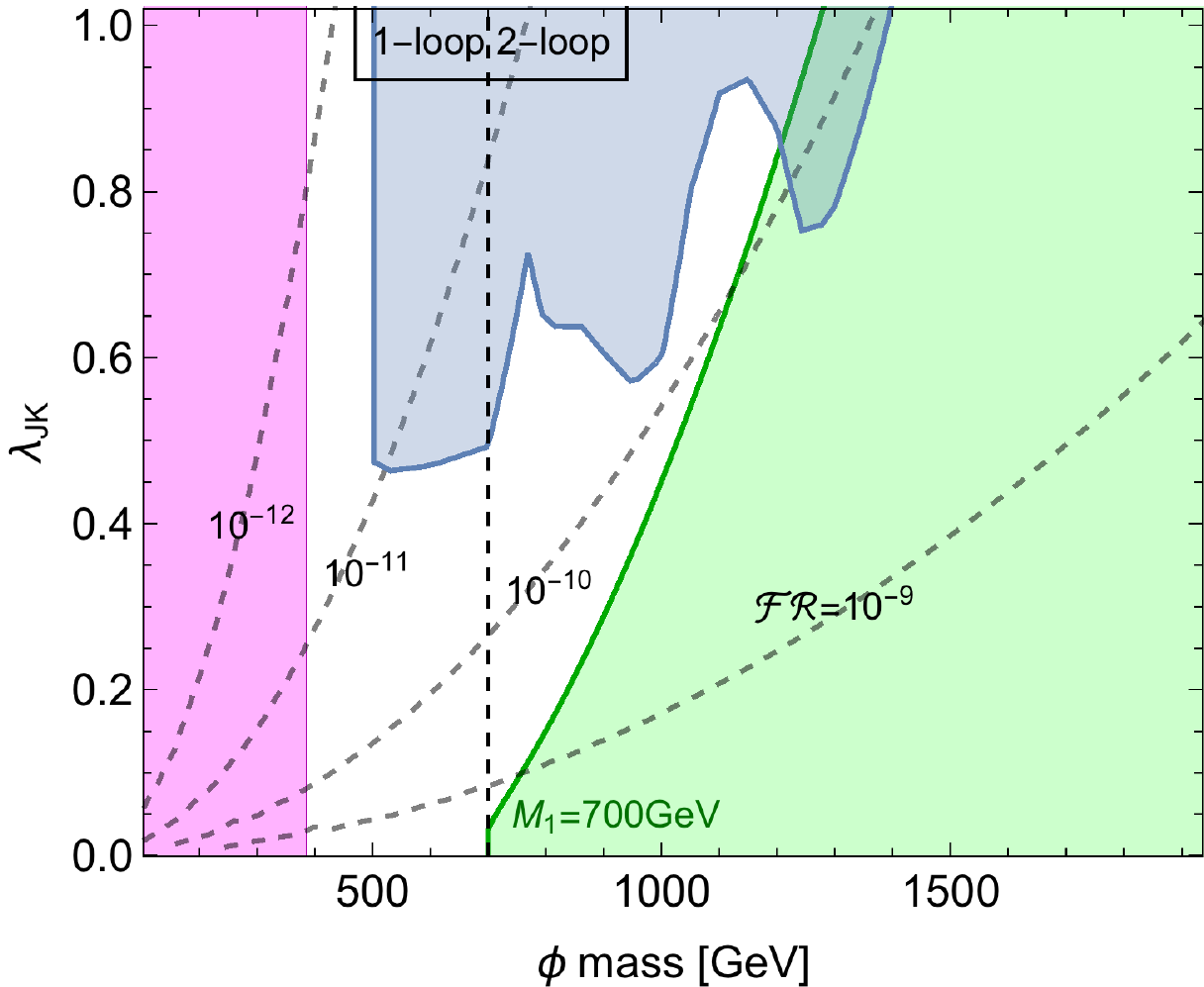} \ 
\includegraphics[height=0.38\textwidth]{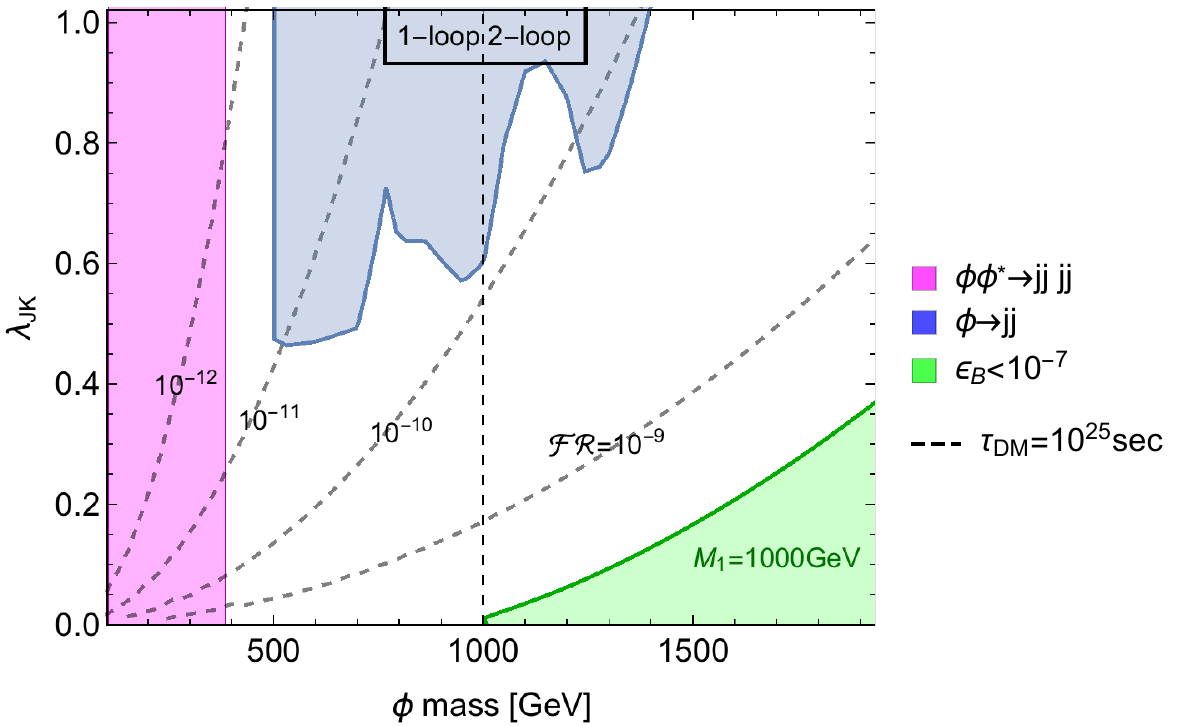} }
\caption{
LHC and dark matter lifetime constraints when a flavor hierarchy is present. The baryon asymmetry is created at one-loop at the left of the dashed line marking $m_\phi = M_1$ and at two-loop on the right. We fix $\kappa_{2;3}=1$, $M_2=2M_1$ and take $M_1=700\gev$ (left) or $M_1=1\tev$ (right). In the green regions the baryon asymmetry is too small. 
Curved dashed lines are contours of $\tau_{DM}=10^{25}\sec$ for the corresponding values of ${\cal F}\cal R$, from Eq.~\eqref{eq:tauDM2loop}. For each value of ${\cal F}\cal R$, the region above the corresponding line is excluded.
The largest coupling $\lambda_{JK}$ involves $q_Jq_K =bs$ for an up-type $\phi$, or $tb$ for a down-type $\phi$. For an up-type $\phi$, the magenta and blue regions are excluded by searches for paired \cite{Khachatryan:2014lpa,ATLASCONF2015026} and single \cite{CMS:2015neg,Aad:2014aqa} dijet resonances, while there are no LHC constraints for down-type scalars. 
}
\label{fig:2looplambda}
\end{center}
\end{figure}

For the case in which $\phi$ is down-type, there are less experimental signatures: resonant production from $\lambda_{33}tb$ would require an initial top quark in the proton, while QCD pair-production can result in $tj$ final states (paired top-jet resonances), $j(tjj)j(\bar tjj)$ (opposite-sign tops + jets) or $jj+\slashed E_T$. Those signatures are  much harder to distinguish from the background and we are not aware of any relevant LHC search. On the other hand,  
typical values of $ {\cal F}\mathcal R$ mentioned earlier are less suppressed than for an up-type $\phi$, so it might be harder to satisfy the DM lifetime in Eq.~\eqref{eq:tauDM2loop}.

We show the combined constraints in Fig.~\ref{fig:2looplambda} by setting $\kappa_{2;3}=1$, $M_1=700\gev$ (left) or $M_1= 1\tev$ (right) and varying $m_\phi$ from $200\gev$ to $2\tev$.
For $m_\phi > M_1 $, the asymmetry is generated at two-loop level as in Eq.~(\ref{epsB2-loop}); we also include the one-loop case of $m_\phi < M_1$  
in order to show the different $\lambda$ dependence between the two cases. 
Shaded regions are excluded by different experimental signatures: at the bottom in green, the asymmetry parameter is too small and cannot generate enough baryons (and dark matter), for each value of $M_1$. Changing the $\Psi_1$ mass gives different phase space suppression factors. Since the twin top is expected to be heavier than $500\gev$ (we have set $m_{\tilde t}=550\gev$), we cannot ignore such suppression, and a relatively large mass of $M_1$ is preferred. 
With dashed lines, we delimit regions excluded by the DM lifetime for typical values of ${\cal F}\mathcal R$ (regions above each curved dashed lines are excluded).
The remaining signatures only apply to an up-type scalar: blue and magenta regions are excluded by the single and $b$-tagged paired dijet searches.  Motivated by flavor arguments, we have assumed that the largest couplings involves $q_Jq_K=bs$. Four-top searches would be able to impose limits only below $m_\phi\simeq 600\gev$, but in this range there are no on-shell $\phi\to t\Psi_1$ decays so there are no exclusions.

At present, we cannot use the current monotop searches without performing a full recasting procedure, which is beyond the scope of this work: in particular the CMS search \cite{Khachatryan:2014uma} does not show limits on $s$-channel produced resonances decaying to top and $\slashed{E}_T$, while the ATLAS search expresses their limits only for a scalar at 500\gev, giving an upper limit on the resonant coupling equal to about $0.1$.  We point out that monotop limits as a function of the $s-$-channel mediator mass would be much more useful. We expect the limit on the monotop cross section to stay approximately constant for higher scalar masses as the signal efficiency would remain the same (or even improve). As the branching ratio ${\rm Br}(\Psi_1\to\tilde q\tilde q\tilde q)$ is easily less than O(10\%) due to phase space suppression from large $m_{\tilde t}$, the monotop limits should be sub-dominant with respect to the ones already presented.

In the allowed parameter space of Fig.~\ref{fig:1loopctau} and Fig.~\ref{fig:2looplambda}, 
we did not include the constraints from $N_{\rm eff}$, 
since it depends on the spectrum of twin light species. 
In  appendix~\ref{DR}, assuming that there is a gauged $U(1)_{\tilde Y}$ in the twin sector and varying the number of light twin neutrinos, we explicitly calculate $\Delta N_{\rm eff}$ for each model, and  overlay the new constraints on Figs.~\ref{fig:1loopctau} and \ref{fig:2looplambda}. The constraints are strong, pointing to heavier masses for the singlets or the need for some model-building to avoid light species in the twin sector.


\section{Conclusions}\label{concl}

In this work, we proposed a simple model for the abundance of dark matter and baryons  
based on Twin Higgs models in the context of asymmetric dark matter scenarios. 
It is promising that this solution of the hierarchy problem naturally yields dark matter masses of the correct magnitude to form an asymmetric dark matter candidate. In particular, dark matter is composed of twin baryons, 
whose number density is equal to that of the baryon as a consequence of a conserved $U(1)_{B-\tilde B}$.  Guided by naturalness, all the relevant masses are expected to be at the (sub)-TeV scale, and  the cosmological/phenomenological consequence of the model are both calculable and within experimental reach. 

In our model, both baryon and twin baryon asymmetries  are simultaneously generated by the decay of a singlet Dirac fermion at a low temperature well below the Twin Higgs cutoff scale.  This Dirac fermion could be either thermally or non-thermally produced. We include both cases and also calculate the corresponding additional relativistic degrees of freedom contributing to $N_{\rm eff}$.  
An essential ingredient for baryogenesis is a new color-charged scalar, and its twin partner. 
Depending on the mass spectrum, the asymmetries are produced at one- or two-loop level.  
Because of separate violation of baryon and twin baryon number, dark matter is metastable and decay to SM hadrons mediated by Dirac fermions and new colored scalars. 
We have discussed the correlations among  
(twin) baryogenesis, dark matter indirect detection signals and collider signatures, especially focusing on the role of the new colored scalars in two benchmark scenarios.
The first interesting example, without a flavor structure for the couplings,
shows a correlation between the cosmological lifetime of dark matter and that of the colored scalar at colliders. 
In order for dark matter to be sufficiently stable, the colored scalar should be long-lived, leaving displaced vertices signatures at LHC. On the other hand, if the colored scalar decays promptly we can expect indirect signals from the dark matter decay. 
In another benchmark point, the baryon asymmetry is generated at two-loop level, so that flavor dependent couplings are needed to generate a sizable baryon abundance and at the same time stabilize the dark matter. Therefore large couplings and lighter spectrum correspond to interesting  collider signatures, among which the resonant production of the colored scalar and monotop events.

In the absence of a light photon and leptons in the twin sector, dark matter could be a mix of twin neutrons and twin protons. As detailed in the appendix, further model-building is necessary to annihilate away the twin pions, e.g. to the SM sector before BBN, or to satisfy constraints on $N_{\rm eff}$. With the twin hypercharge present, further investigating dark sector phenomenology such as dark nucleosynthesis or dark matter self-interactions would be an interesting direction for future work.

Our mechanism could also be applied to various iterations of the TH paradigm, for example Fraternal TH, or even to models with unconfined hidden non-abelian gauge groups. However, in this case it is not clear how to naturally obtain a $5\gev$ dark matter mass, or a sizable annihilation cross-section that only leaves the asymmetric component of dark matter.

\small

\section*{Acknowledgements}
This work is supported by DOE grant
DOE-SC0010008.

\appendix

\section{Twin chiral Lagrangian}\label{app:chiral}
Here we discuss the long-distance hadronic matrix elements for the twin neutron operator in Eq.~(\ref{eq:qqqO}). For this we develop a chiral Lagrangian for light twin quarks, following the procedure to derive the SM chiral Lagrangian as in Refs.~\cite{Claudson:1981gh,Aoki:1999tw,Davoudiasl:2011fj}.
We will assume that, in similarity to the SM, the twin sector has three light twin quarks, such that the twin chiral transformation group is $SU(3)_{\tilde L}\times SU(3)_{\tilde R}$. The pseudo-Goldstone bosons associated with spontaneous chiral symmetry breaking can be expressed by a $3\times 3$ special unitary matrix, $\Sigma=\exp(2i\tilde M/f_{\tilde\pi})$. Twin mesons and baryons are denoted by $3\times3$ matrices,
\beq
 \tilde M=\left(\begin{array}{ccc} \frac{\tilde \eta^0}{\sqrt6}+\frac{\tilde \pi^0}{\sqrt2} & \tilde \pi^+  & \tilde K^+  \\  \tilde \pi^-& \frac{\tilde \eta^0}{\sqrt6}-\frac{\tilde \pi^0}{\sqrt2}   & \tilde K^0  \\\tilde K^-  & \tilde{\bar K}^0  & -\sqrt{\frac{2}{3}}\tilde \eta^0  \end{array}\right)\,,\qquad 
\tilde B=\left(\begin{array}{ccc} \frac{\tilde \Lambda^0}{\sqrt6}+\frac{\tilde \Sigma^0}{\sqrt2} & \tilde \Sigma^+  & \tilde p  \\  \tilde \Sigma^-& \frac{\tilde \Lambda^0}{\sqrt6}-\frac{\tilde \Sigma^0}{\sqrt2}   & \tilde n  \\\tilde \Xi^-  & \tilde{\Xi}^0  & -\sqrt{\frac{2}{3}}\tilde \Lambda^0  \end{array}\right)\,.
\eeq
Under the $SU(3)_{\tilde L}\times SU(3)_{\tilde R}$ symmetry, those transform as $\Sigma\to \tilde L\Sigma \tilde R^\dag$ and $B\to U B U^\dag$, where $U$ are non-linear functions of $\tilde L,\tilde R,\tilde M$ defined by the transformation properties of $\xi=\exp (i{\tilde M}/{ f_{\tilde\pi}})$, that is, $\xi\to \tilde L\xi  U^\dag= U\xi \tilde R^\dag$.

At lowest-order, the $\tilde B$-conserving Lagrangian invariant under $SU(3)_{\tilde L}\times SU(3)_{\tilde R}$ is
\beq
{\cal L}_0&=\frac{f_{\tilde\pi}^2}{8}\tr(\partial^\mu\Sigma)(i\partial_\mu\Sigma^\dag)+\tr \tilde{\bar B}(i\gamma^\mu\partial_\mu-M_{\tilde B})\tilde{B}\nn\\
& +\frac{1}{2}i\tr\tilde{\bar B}\gamma^\mu\left[\xi\partial_\mu\xi^\dag
+\xi^\dagger\partial_\mu\xi\right]\tilde B 
+\frac{1}{2}i\tr\tilde{\bar B}\gamma_\mu \tilde B\left[(\partial_\mu\xi)\xi^\dag
+(\partial_\mu\xi^\dag)\xi\right] \nn \\
& -\frac{1}{2}i(\tilde D-\tilde F)\tr\tilde{\bar B}\gamma^\mu\gamma_5 \tilde B \left[(\partial_\mu\xi)\xi^\dag-(\partial_\mu\xi^\dag)\xi\right] \nn \\
& +\frac{1}{2}i(\tilde D+\tilde F)\tr\tilde{\bar B}\gamma^\mu\gamma_5\left[\xi\partial_\mu\xi^\dag-\xi^\dagger\partial_\mu\xi\right]\tilde B.
\label{eq:chiral0}
\eeq
For SM quarks, the parameters $D$ and $F$  for the axial-vector matrix elements can be extracted from semi-leptonic baryon decays, which gives $\tilde D=0.80$ and $\tilde F=0.47$~\cite{Cabibbo:2003cu};
we expect those to be the same in the twin sector. $f_{\tilde\pi}$ is the twin pion decay constant, and is related to the SM quantity by $f_{\tilde\pi}=f_\pi \tilde \Lambda_{QCD}/\Lambda_{QCD}$; the above normalization of the kinetic terms corresponds to $f_\pi=130$ MeV.
 Twin quark masses are not invariant and a symmetry-breaking Lagrangian ${\cal L}_1$ can be written, with the masses as spurions; these effects are sub-leading in this work.

We now want to find the chiral Lagrangian for $\tilde B$-violating interactions: microscopically, they arise from the Lagrangian in Eq.~\eqref{eq:DMlagr}, which only has right-handed twin quarks $\tilde u^c,\tilde d^c,\tilde s^c$. We can write the terms relevant for DM decay as ${\cal L}_{\tilde{\slashed B}}\supset \sum_i c_i  (\tilde O_iq)(qq)$, where we defined the operators $\tilde O_i$,
\beq
\tilde O_1&=\epsilon_{ \alpha \beta \gamma}
(\tilde d^{\alpha}\tilde s^{\beta})\tilde u^{\gamma} \,,\quad 
\tilde O_2=\epsilon_{ \alpha \beta \gamma}
(\tilde u^{\alpha}\tilde d^{\beta})\tilde d^{\gamma} \,,\\
\tilde O_3&=\epsilon_{ \alpha \beta \gamma}
(\tilde s^{\alpha}\tilde u^{\beta})\tilde d^{\tilde\gamma} \,,\quad
\tilde O_4=\epsilon_{ \alpha \beta \gamma}
(\tilde u^{\alpha}\tilde d^{\beta})\tilde s^{\gamma} \,.
\eeq
Here, the spinor indices are contracted within each parenthesis and $\alpha,\beta,\gamma$ are twin color indices. For now we drop all SM flavor indices. The coefficients $c_i$ have dimension of mass$^{-5}$ and non-trivial flavor structures from the couplings in Eq.~\eqref{eq:DMlagr}. In our microscopic model, it can easily be recognized that the first operator comes from an up-type $\tilde \phi$ exchange,  with $c_1\propto \tilde \kappa_{A;1}\tilde\lambda_{12}$, while the remaining ones correspond to the down-type $\tilde \phi$ scenario,  $c_2\propto \tilde \kappa_{A;1}\tilde\lambda_{11}, c_3\propto~\tilde \kappa_{A;1}\tilde\lambda_{12}, \ c_4\propto \tilde \kappa_{A;2}\tilde \lambda_{11}$.\footnote{At loop level, twin EW processes give additional mixing contributions, denoted by $\FO$ in the text.} Only three operators are independent, as $\tilde O_1+\tilde O_3+\tilde O_4=0$ via a Fierz identity: for example, we could have eliminated $\tilde O_4$ and absorbed the coefficient $c_4$ into  $c_1$ and $c_3$, but the dependence on the UV couplings would be less intuitive.

The Lagrangian can be rewritten as ${\cal L}_{\tilde{\slashed B}}=(
{\rm Tr}[{\cal C} \tilde O]q)(qq)$, with
\beq
{\cal C}=\left(\begin{array}{ccc}c_1& 0 & 0 \\0 & c_3 & 0 \\0 & c_2 & c_4\end{array}\right),\qquad
 \tilde O=
\left(\begin{array}{ccc}\tilde O_1& 0 & 0 \\0 & \tilde O_3 & \tilde O_2 \\0 & 0 & \tilde O_4\end{array}\right)\,.
\eeq
The right-handed light twin quarks ($\tilde u^c,\tilde d^c,\tilde s^c$) transform in the  ({\bf1},{\bf 3}) representation of the chiral symmetry, such that the operators $ \tilde O_i$ transform as $({\bf1},{\bf8})$. ${\cal C}$ can thus be treated as a spurion transforming as $({\bf1},{\bf8})$ to make ${\rm Tr}[{\cal C}\tilde O]$ is invariant.
Finally, the chiral $ \tilde B$-violating Lagrangian will be given by invariant combinations of meson and baryon fields containing the spurion ${\cal C}$:
\beq\label{eq:twinchiral}
{\cal L}_{\tilde{\slashed B}}=\tilde \beta ({\rm Tr}[{\cal C} \xi^\dag \tilde B\xi] q)(qq)\,+h.c.
\eeq
For now, $\tilde \beta$ is an unknown prefactor of dimension mass$^3$.
Expanding the Lagrangian up to first order in $1/f_{\tilde\pi}$ and only keeping terms with the twin neutron, we find the terms relevant for dark matter decays:
\beq\label{eq:DMchiral}
&{\cal L}_{\tilde n}\supset\tilde \beta c_2 (\tilde nq)(qq) \ +\  i\frac{\tilde \beta}{  f_{\tilde\pi}} \left[\left(\frac{\tilde \pi ^0}{\sqrt{2}}-\sqrt{\frac{3}{2}}\tilde \eta^0\right)c_2+( c_3-c_4)  \tilde {\bar K}^0\right] (\tilde nq)(qq)
\eeq

\begin{figure}[t]
\begin{center}
\includegraphics{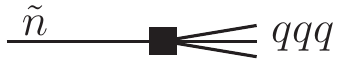}\hfill
\includegraphics{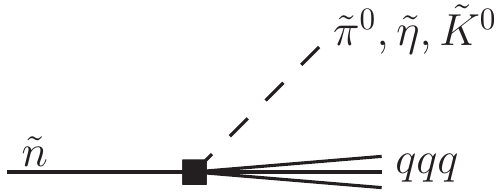}\hfill
\includegraphics{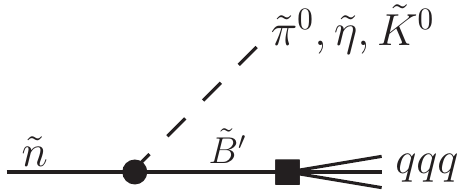}
\caption{Diagrams for twin neutron decays to three SM quarks and up to one twin meson. Square vertices indicate $\tilde B$-violating interactions in Eq.~\eqref{eq:DMchiral}, while the round vertex is a $\tilde B$-conserving interaction with a twin meson from Eq.~\eqref{eq:chiral0}. In the last diagram $\tilde B'$ indicates an intermediate twin baryon, e.g. $\tilde B'=\tilde \Lambda^0$. }
\label{fig:chiraldecay}
\end{center}
\end{figure}

The first term describes twin neutron decay to three SM quarks, while the other terms produce an additional twin meson. From this $\tilde B$-violating interactions, Eq.~(\ref{eq:DMchiral}),  and  the $\tilde B$-conserving terms, Eq.~(\ref{eq:chiral0}), 
the twin neutron decay can be computed via the diagrams in Fig.~\ref{fig:chiraldecay}. 
The corresponding matrix elements between the twin neutron and up to one twin meson 
are given by 
\beq\label{eq:twinME}
\langle 0 |  \tilde d^c(\tilde u^c \tilde d^c) |\tilde n \rangle =&\tilde\beta,
\nn\\
\langle \tilde \eta^0|\tilde d^c(\tilde u^c\tilde d^c)|\tilde n\rangle \simeq & 
\frac{\tilde\beta  }{f_{\tilde\pi}} 
 \left(-\sqrt{\frac{3}{2}}+\frac{(\tilde D - 3\tilde F)}{\sqrt{6}}\right),\nonumber\\
\langle \tilde\pi^0|\tilde d^c(\tilde u^c\tilde d^c )  |\tilde n\rangle \simeq & \frac{\tilde \beta  }{ f_{\tilde\pi}} 
\left(\frac{1}{\sqrt{2}}+  \frac{\tilde D+\tilde F}{\sqrt{2}}\right),
 \\
 \langle \tilde K^0|\tilde s^c(\tilde u^c\tilde  d^c) |\tilde n\rangle \simeq & \frac{\tilde\beta  }{ f_{\tilde\pi}} 
 \left(-1 - \frac{(\tilde D + 3 \tilde F)}{3}\frac{m_{\tilde n}}{m_{\tilde \Lambda}}\right),
  \nn\\
 \langle \tilde K^0 |\tilde u^c(\tilde d^c\tilde  s^c) |\tilde n\rangle \simeq &
 \frac{\tilde \beta}{ f_{\tilde\pi}}
  \left(\frac{ 2 \tilde D }{3}\frac{m_{\tilde n}}{m_{\tilde \Lambda}}\right),\nn
\eeq
In the SM, the factor  $\beta\equiv\langle 0|d^c(u^cd^c)|n\rangle$ can be computed via lattice QCD \cite{Aoki:1999tw}, giving $\beta=0.014\gev^3$ $\simeq \Lambda_{QCD}^3$. 
The matrix elements with one twin meson are given at leading order in a $q^2/m_{\tilde n}^2$ expansion, 
where $q$ is the total momentum of the SM decay products;
higher order corrections in the $q^2$ expansion are expected to give $O(10\%)$ corrections, as $m_N^2\lesssim q^2<m_{\tilde n}^2$.
A full list of matrix elements for SM baryon decays can be found in Ref.~\cite{Aoki:1999tw}, and can easily be translated into  twin matrix elements.

As already discussed,  in comparison with the value of the SM counterparts, 
we expect dimensionless couplings ($D,\, F$) to stay the same, while dimensionful ones $(\beta,\, f_\pi, \, m_N,\, m_\Lambda)$ will scale by appropriate powers of $\tilde \Lambda_{QCD}/\Lambda_{QCD}\simeq 5$:
\beq
\tilde\beta \simeq \tilde \Lambda_{QCD}^3,\  
 f_{\tilde\pi} \simeq \frac{\tilde \Lambda_{QCD}}{\Lambda_{QCD}}f_\pi,\ \tilde D\simeq 0.80,\ \tilde F\simeq  0.46,\, 
\ \frac{ m_{\tilde n}}{m_{\tilde \Lambda}}=\frac{ m_{ n}}{m_{\Lambda}}\simeq 0.82\, .
\eeq

\section{Dark radiation}\label{DR}
In a complete Twin Higgs model, additional light degrees of freedom can arise from twin neutrinos and the twin photon. In addition, in our asymmetric dark matter model, one needs to make sure that the symmetric twin component has annihilated away. Without either a twin photon or a $\bar qq\bar{\tilde q}\tilq$ portal, the twin pion would be stable and would have a large abundance $Y_{B+\tilde B}\sim Y_{\Psi_1}\gg Y_{B-\tilde B}$. If a twin photon is present, the decay $\tilde \pi^0\to\tilde \gamma\tilde\gamma$ is allowed, while a pion portal allows the release of the twin entropy back into the SM sector, in which case the decay must happen before BBN, that is, $\tau_{\tilde \pi}<1\sec$.

Dark radiation (DR) is an important constraint on twin Higgs models when we introduce a 
twin $U(1)_Y$ symmetry,  even though the explicit constraint depends on the detailed particle content. 
Compared to the ordinary twin Higgs,  in our baryogenesis scenario there is one more source of dark radiation, the out-of-equilibrium decays of the singlet $\Psi_1$. 
If $\Psi_1$ decays before the two sectors are thermally decoupled, the dark radiation abundance is not different from the case of normal twin Higgs model. 
Then, 
\beq
\left(\frac{\rho_{\widetilde{SM}}}{\rho_{SM}}\right)_{T_{\rm dec}} = \frac{\tilde g_* (T_{\rm dec})}{g_*(T_{\rm dec})}, 
\eeq where $\tilde g_*$ is the effective massless degrees of freedom for the twin sector. 
The final dark radiation abundance  represented by $\Delta N_{\rm eff}$ is 
\beq
\Delta N_{\rm eff}= \left(\frac{\tilde\rho_{\rm rad}}{\rho_{\nu}}\right)_{T_{BBN}}=
\left(\frac{\tilde g_*(T_{BBN})}{g_\nu}\right)
\left(\frac{g_{*S}(T_{BBN})}{\tilde g_{*S}(T_{BBN})}\right)^{4/3} 
\left(\frac{\tilde g_{*S}(T_{\rm dec})}{g_{*S}(T_{\rm dec})}\right)^{4/3},
\eeq
where $T_{BBN}$ is the temperature well below the electron mass, at which light nuclei are generated, $g_\nu$ is the effective degrees of freedom for a single light neutrino, $g_\nu=(7/4)(4/11)^{4/3}\simeq 0.45$.  
For the SM, $g_{*S}\simeq 3.9$ at $T< m_e$. For the twin sector, with an approximated $Z_2$ symmetry, it is reasonable to assume that the light twin neutrinos are decoupled before twin electron-positron annihilation. Then, 
\beq
\Delta N_{\rm eff} =&\,13.68 \left(\frac{2+   N_{\tilde \nu} (7/4)(4/11)^{4/3}}{(2+ N_{\tilde \nu} (7/4)(4/11))^{4/3}}\right) 
\left(\frac{\tilde g_{*S}(T_{\rm dec})}{g_{*S}(T_{\rm dec})}\right)^{4/3}, 
\eeq where $N_{\tilde \nu} $ is the number of light (left-handed) twin neutrinos. 
This is potentially large. The optimistic example is that 
the two sectors  decouple at $\Lambda_{QCD} < T_{\rm dec} < \widetilde\Lambda_{QCD}$.  
 At this temperature, 
$g_{*S}(T_{\rm dec}) \simeq 61.75$, and $\tilde g_{*S}(T_{\rm dec}) = 2+ (N_{\tilde \nu}  + 2)(7/4)$. 
The corresponding $\Delta N_{\rm eff}$ is
\beq
\Delta N_{\rm eff} 
= &\, 0.056\left(\frac{(2+ N_{\tilde \nu}  (7/4)(4/11)^{4/3})(2+ (N_{\tilde \nu}  + 2)(7/4))^{4/3}
}{(2+ \tilde N_\nu (7/4)(4/11))^{4/3}} \right)\nonumber\\
=&\, (0.43,\, 0.53,\,0.63,\,0.73)\quad{\rm for}\ N_{\tilde \nu}  =(0,\,1,\,2,\,3).
\eeq

One the other hand, if $\Psi_1$ decays after decoupling, the twin sector daughter particles also contribute to the dark radiation. Here we consider a case that this energy density $\Delta \rho_{\widetilde{SM}}$ is greater than 
the background twin radiations as $\Delta \rho_{\widetilde{SM}} = {\rm Br}_{\Psi_1\to \widetilde{SM}}\, M_1 n_{\Psi_1}\ \simeq  (\pi^2 \tilde g(\tilde T_D)/30) \tilde T_D^4$, where the twin sector is quickly thermalized at a temperature $\tilde T_D$. 
We get
 \beq
\left(\frac{\tilde\rho_{SM} }{\rho_{SM}}\right)_{T_D}\simeq 
{\rm Br}_{\Psi_1\to\tilde q\tilde q\tilde q} \left(\frac{4 g_{*S}(T_D)}{3 g_*(T_D)}\right) \left(\frac{M_1}{T_D}\right) Y_{\Psi_1}, 
\eeq and 
\beq
\Delta N_{\rm eff} = &\,
\left(\frac{\tilde g_*(\tilde T_{BBN})}{g_\nu}\right)
\left(\frac{g_{*S}(T_{BBN})}{\tilde g_{*S}(\tilde T_{BBN})}\right)^{4/3} 
 \left(\frac{\tilde g_{*S}(\tilde T_D)}{g_{*S}(T_D)}\right)^{4/3}\nonumber\\
&\,\times \left(\frac{4}{3}{\rm Br}_{\Psi_1\to\widetilde{SM}}\right)
\left(\frac{g_{*S}(T_D)}{\tilde g_*(\tilde T_D)}\right)
\left( \frac{M_1}{T_D} \right)
\left(\frac{Y_{\Delta B}}{\epsilon}\right), 
\eeq
where $T_D$ is the temperature of the SM at the time of $\Psi_1$ decays,
 and it should be greater than $\tilde T_D$, $\tilde T_{BBN}$ is the twin sector temperature at $T= T_{BBN}$ for the SM.
Assuming that $\tilde T_D$ is higher than the light twin neutrino decoupling temperature,
\beq
\Delta N_{\rm eff} = &\, 0.46\,{\rm Br}_{\Psi_1\to \widetilde{SM}} 
\left(\frac{(2 + N_{\tilde \nu}  (7/4)(4/11)^{4/3})(2 + (N_{\tilde \nu}  + 2)(7/4))^{1/3}}{(2 +N_{\tilde \nu} (7/4)(4/11))^{4/3}}\right)
\nonumber\\
&\,\times \left(\frac{M_1}{10^3 T_D}\right)\left(\frac{10^{-6}}{\epsilon}\right)\left(\frac{60}{g_*(T_D)}\right)^{1/3}
\\
=&\, (0.65,\, 0.61,\,  0.58,\, 0.56)\,{\rm Br}_{\Psi_1\to \widetilde{SM}}
\left(\frac{M_1}{10^3 T_D}\right)\left(\frac{10^{-6}}{\epsilon}\right)
\left(\frac{60}{g_*(T_D)}\right)^{1/3} 
\quad {\rm for}\ N_{\tilde \nu}  = (0,\, 1,\, 2,\, 3).\nonumber
\eeq

In the case of non-thermal production of $\Psi_1$ with a low reheating temperature, 
all the SM, twin SM radiations and singlets are produced by the decay of the reheaton, $\phi_{\rm reh}$.  
The twin radiation energy density produced  by direct decay of the reheaton could be suppressed by small branching fractions as ${\rm Br}_{\phi_{\rm reh}\to
\widetilde {SM}}\ll {\rm Br}_{\phi_{\rm reh}\to SM}$ . However, that from the singlet decay might not be suppressed because of constraint from baryon asymmetry. 
$  \rho_{\widetilde{SM}}$ from $\Psi_1$ decays is \beq
\left(\frac{\rho_{\widetilde{SM}}}{\rho_{SM}}\right)_{T_{\rm reh}} =
{\rm Br}_{\Psi_1\to\widetilde {SM}} \left(\frac{{\rm Br}_{\phi_{\rm reh}\to \Psi_1}}{{\rm Br}_{\phi_{\rm reh}\to SM}} \right) \simeq
{\rm Br}_{\Psi_1\to\widetilde {SM}} \left(\frac{4 g_{*S}(T_{\rm reh})}{3 g_*(T_{\rm reh})}\right) \left(\frac{m_{\rm reh}}{T_{\rm reh}}\right) Y_{\Psi_1}.
\eeq
$m_{\rm reh}$ is the mass of the reheaton. Compared to the above formula of $\Delta N_{\rm eff}$, 
 $T_D$ and $M_1$ should be changed to  $T_{\rm reh}$ and $m_{\rm reh}$, respectively.
 That is  
\beq
\Delta N_{\rm eff} = &\, 0.46\, {\rm Br}_{\Psi_1\to \widetilde{SM}} 
\left(\frac{(2 + N_{\tilde \nu}  (7/4)(4/11)^{4/3})(2 + (N_{\tilde \nu}  + 2)(7/4))^{1/3}}{(2 + N_{\tilde \nu} (7/4)(4/11))^{4/3}}\right)
\nonumber\\
&\,\times \left(\frac{m_{\rm reh}}{10^3 T_{\rm reh}}\right)\left(\frac{10^{-6}}{\epsilon}\right)\left(\frac{60}{g_*(T_{\rm reh})}\right)^{1/3}
\label{eq:DeltaNeffNTH}
\\
=&\, (0.65,\, 0.61,\,  0.58,\, 0.56)\, {\rm Br}_{\Psi_1\to \widetilde{SM}}
 \left(\frac{m_{\rm reh}}{10^3 T_{\rm reh}}\right)\left(\frac{10^{-6}}{\epsilon}\right)\left(\frac{60}{g_*(T_{\rm reh})}\right)^{1/3}
\quad {\rm for}\ N_{\tilde \nu} = (0,\, 1,\, 2,\, 3).\nonumber
\eeq

\begin{figure}[tbhp]
\begin{center}
\includegraphics{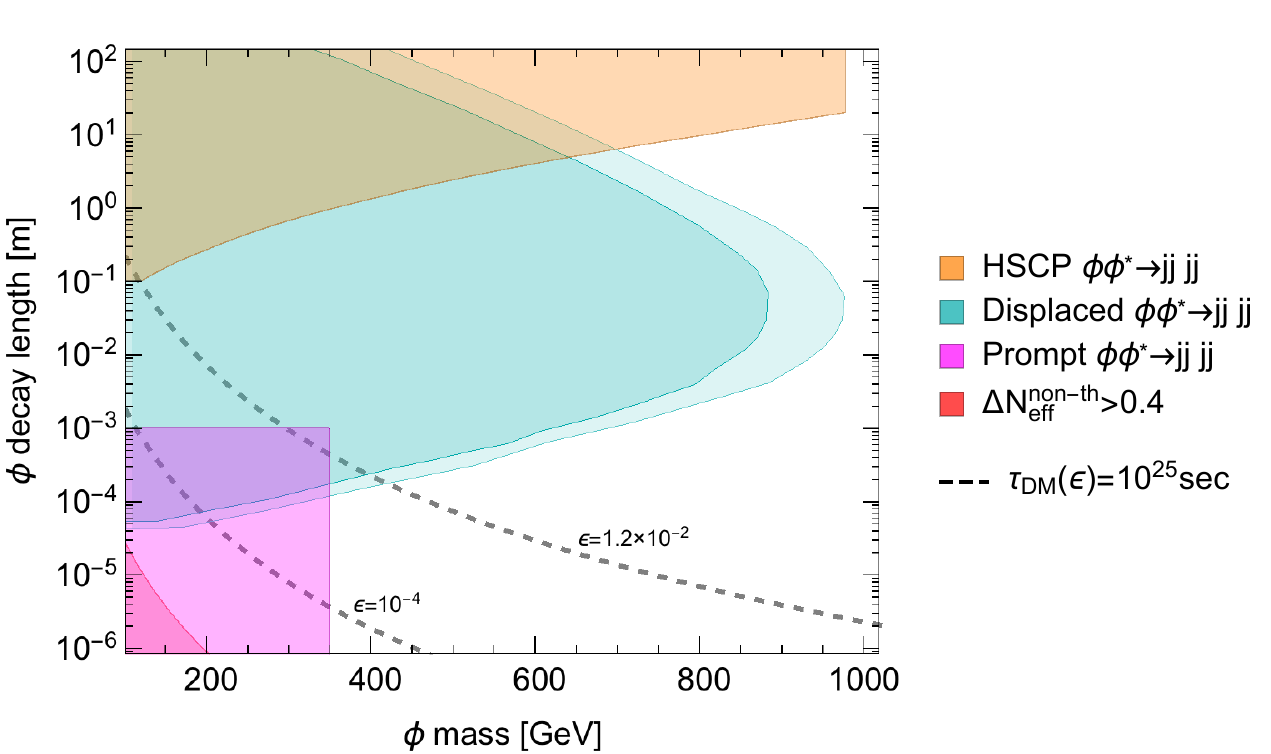}
\caption{Same as Fig.~\ref{fig:1loopctau}, but including the constraints from $\Delta N_{\rm eff}$ (in red) on the one-loop model. For consistency with Fig.~\ref{fig:1loopctau}, we showed the constraints from non-thermal $\Psi_1$ production, Eq.~\eqref{eq:DeltaNeffNTH}.
}
\label{fig:neff1}
\end{center}
\end{figure}

\begin{figure}[tbhp]
\begin{center}
\makebox[\textwidth][c]{
\includegraphics[height=0.38\textwidth]{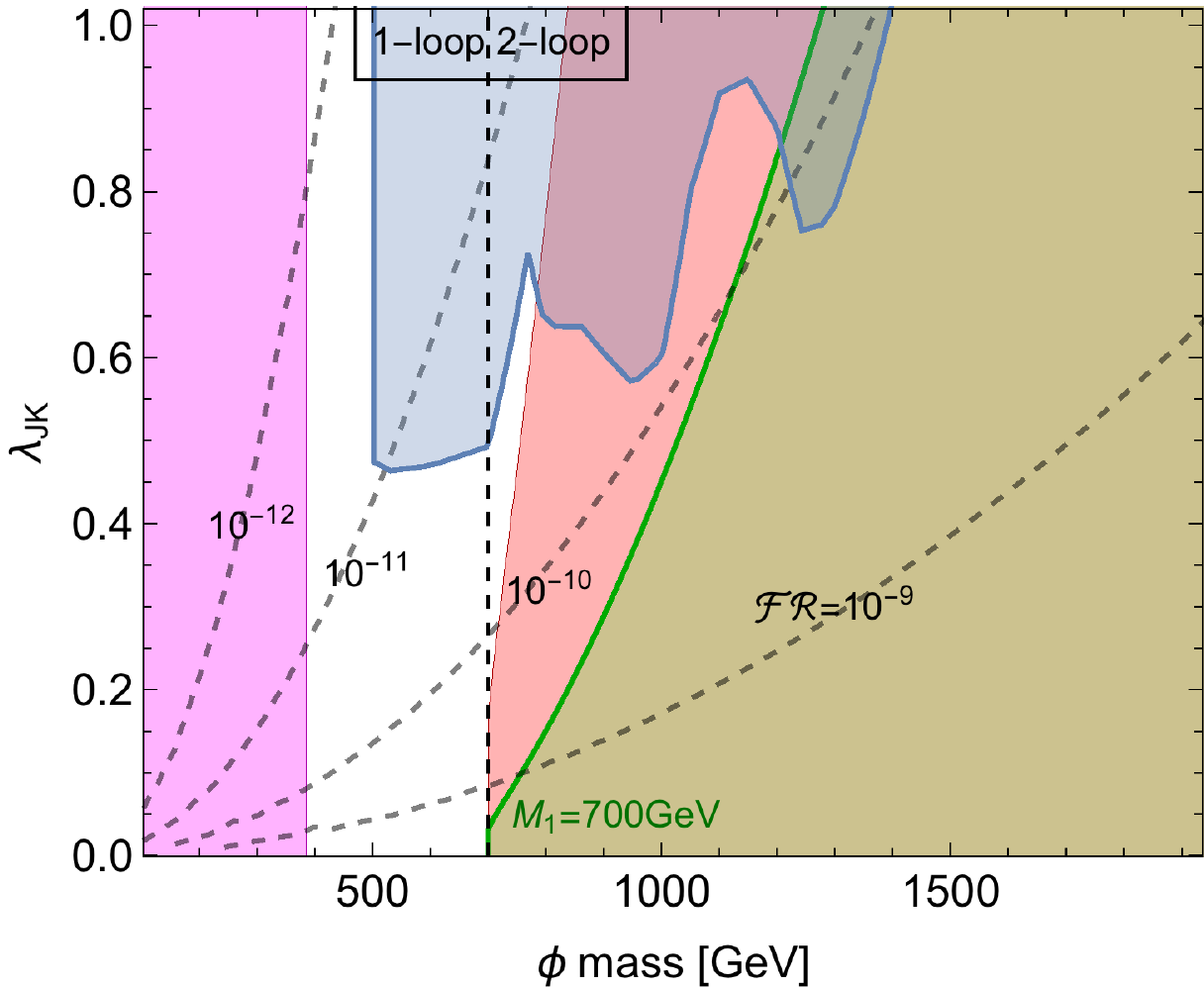} \ 
\includegraphics[height=0.38\textwidth]{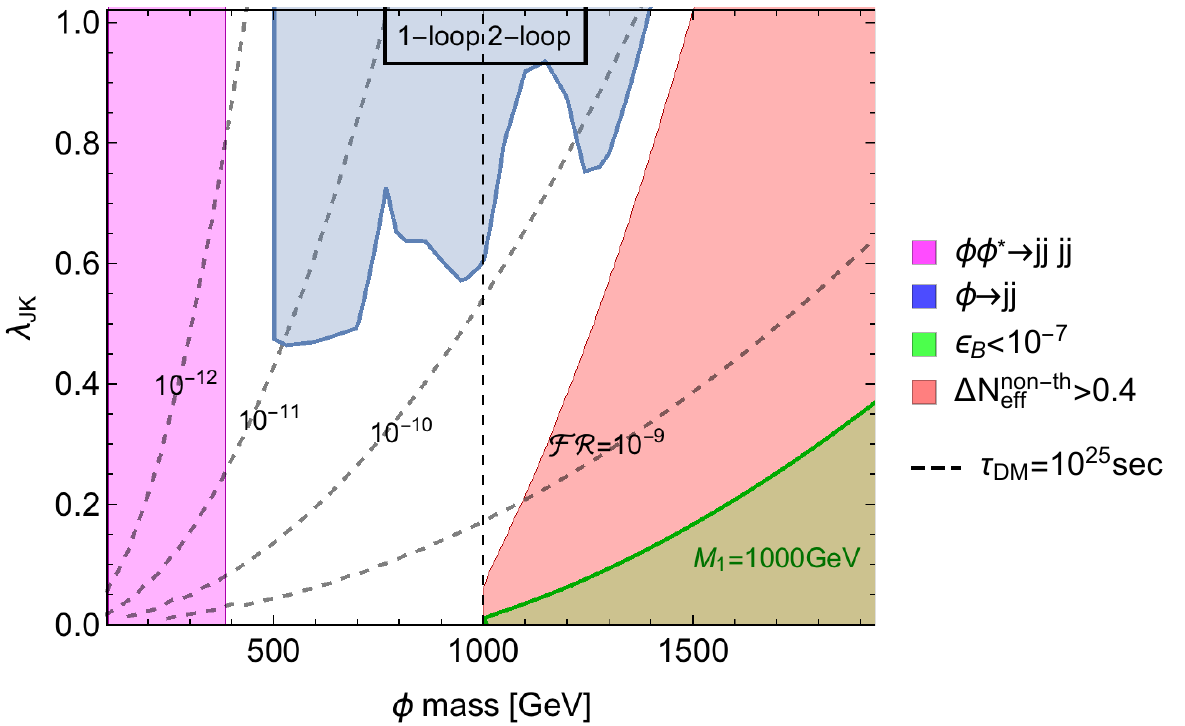}
}
\caption{Same as Fig.~\ref{fig:2looplambda}, but including the constraints from $\Delta N_{\rm eff}$ (in red) on the two-loop model with hierarchical couplings.
}
\label{fig:neff2}
\end{center}
\end{figure}

These values should be compared to the experimental limit from Planck 2015 \cite{Ade:2015xua},
$|\Delta N_{\rm eff}|<0.18$ (68\%CL).\footnote{A
recent study \cite{Riess:2016jrr} reduced the uncertainties on  the local Hubble constant $H_0$ 
using the Hubble Space Telescope, based on updated distance measurements of Cepheid variables 
in galaxies with recent type-Ia supernovae. 
The measured value of $H_0$ shows a $3\,\sigma$ difference from that of Planck. 
The tension could be resolved if there is additional dark radiation with $\Delta N_{\rm eff} = 0.4-1$, which can naturally be reproduced in our model. In the following, we show the constraint, $\Delta N_{\rm eff} < 0.4$ (corresponding to Planck's $95\%CL$ interval) in the plots. Limits from $\Delta N_{\rm eff}\leq1$ would be just slightly weaker.}
Note that for large $Y_{\Psi_1}$  (and as a result large $\Delta N_{\rm eff}$), the asymmetry parameter is smaller to reproduce the observed baryon abundance, so that the $\Delta N_{\rm eff}$ bound roughly translates into $\epsilon>10^{-6}$; therefore, the two-loop baryogenesis model would be highly constrained by dark radiation. 
We show the resulting additional constraint (red color) on the one-loop and two-loop models in Figs.~\ref{fig:neff1}-\ref{fig:neff2}.


\bibliography{refs}

\end{document}